\documentclass[submission,copyright,creativecommons]{eptcs}
\providecommand{\event}{MARS 2022} 
\usepackage{graphicx}
\usepackage{soul} % to highlight text
\usepackage{multirow}
\usepackage{cite}
\usepackage{xspace}
\usepackage[caption=false]{subfig}
\usepackage[latin1]{inputenc}
\usepackage{comment}
\usepackage{tikz}
\usetikzlibrary{automata,positioning}
\usepackage{pifont}
\usepackage{listings}
\usepackage{xcolor}
\usepackage{cadp-lnt}
\usepackage{cadp-mcl}
\colorlet{punct}{red!60!black}
\definecolor{background}{HTML}{F9F9F8}
\definecolor{delim}{RGB}{20,105,176}
\usetikzlibrary{arrows.meta,
                chains,
                decorations.pathmorphing,
                positioning,
                shapes.geometric,
                shapes.symbols
                }

\definecolor{mygreen}{rgb}{0,0.6,0}
\definecolor{mygray}{rgb}{0.5,0.5,0.5}
\definecolor{mymauve}{rgb}{0.1,0.2,0.7}
\definecolor{olivegreen}{cmyk}{.6,.4,0.8,0}
                
\makeatletter
\def\mylabel#1#2{\@bsphack\if@filesw {\let\thepage\relax
   \def\protect{\noexpand\noexpand\noexpand}%
   \edef\@tempa{\write\@auxout{\string
      \newlabel{#1}{{#2}{\thepage}{}{figure.1.1}{}}}}%
   \expandafter}\@tempa
   \if@nobreak \ifvmode\nobreak\fi\fi\fi\@esphack}
\makeatother

\title{Formally Modeling Autonomous Vehicles in LNT for Simulation and Testing}

\author{Lina Marsso 
\institute{\textit{Dept. of Computer Science} \\
\textit{University of Toronto}\\
Toronto, Canada \\}
\email{lina.marsso@utoronto.ca}
\and Radu Mateescu
\institute{\textit{Univ.~Grenoble Alpes}\\
\textit{Inria, CNRS, Grenoble~INP\thanks{Institute of Engineering Univ.~Grenoble Alpes}, LIG}\\
38000 Grenoble, France  \\}
\email{radu.mateescu@inria.fr}
\and Lucie Muller
\institute{\textit{Univ.~Grenoble Alpes}\\
\textit{Inria, CNRS, Grenoble~INP${}^*$, LIG}\\
38000 Grenoble, France  \\}
\email{lucie.muller@inria.fr}
\and Wendelin Serwe
\institute{\textit{Univ.~Grenoble Alpes}\\
\textit{Inria, CNRS, Grenoble~INP${}^*$, LIG}\\
38000 Grenoble, France  \\}
\email{wendelin.serwe@inria.fr}}

\def\titlerunning{Formally Modeling Autonomous Vehicles in LNT for Simulation and Testing}
\def\authorrunning{L. Marsso, R. Mateescu, L. Muller, and W. Serwe}

\begin{document}

\maketitle              % typeset the header of the contribution

\begin{abstract}
We present two behavioral models of an autonomous vehicle and its interaction with the environment.
Both models use the formal modeling language LNT provided by the CADP toolbox.
This paper discusses the modeling choices and the challenges of our autonomous vehicle models, and also illustrates how formal validation tools can be applied to a single component or the overall vehicle.
\end{abstract}

\section{Introduction}

Autonomous vehicles (AV) are complex safety critical systems, as undesired behaviours can lead to fatal accidents~\cite{boudette-21}.
Both the complete AV and its components need to be tested to handle critical scenarios.
Because these critical scenarios are unlikely to happen in real environments, a common practice in robotics~\cite{UNECE-21} is to reproduce these critical scenarios in an autonomous driving (AD) simulator, such as CARLA~\cite{Dosovitskiy17}.
The specifications of these critical scenarios are obtained manually, by random generation, or by derivation from a (formal) model~\cite{Fremont-Dreossi-Ghosh-et-al-19,Riedmaier-et-al-20,Horel-Laugier-Marsso-Mateescu-Muller-Paigwar-Renzaglia-Serwe-22}.

The main contributions of this paper are
(a) two \emph{formal} models of an AV and its environment, which can be used to generate relevant critical scenarios for testing AVs and/or their components, and
(b) a discussion comparing the models and motivating the existence of two different models by their intended principal uses.
Both models describe an autonomous ego vehicle, called car, moving around in a scene, called (geographical) map, towards a goal or destination position, and interacting with its environment, i.e., a given set of moving obstacles (pedestrians, cyclists, other cars, etc.) to avoid collisions.
The formal models are written in the LNT language~\cite{Garavel-Lang-Serwe-17,Champelovier-Clerc-Garavel-et-al-10-v7.0}, which is the most recent modeling language supported by the CADP verification toolbox~\cite{Garavel-Lang-Mateescu-Serwe-13} and a state-of-the-art replacement for the international standards LOTOS and E-LOTOS~\cite{Garavel-Lang-Serwe-17}.
We chose LNT rather than a scenario modeling language such as Scenic~\cite{Fremont-Dreossi-Ghosh-et-al-19} to illustrate a model-based, formal verification and testing approach~\cite{Marsso-Mateescu-Serwe-20,Horel-Laugier-Marsso-Mateescu-Muller-Paigwar-Renzaglia-Serwe-22}.
The first formal model specifies the control of the car (including route planning), considering an abstract representation of the geographical map as a graph.
The second model focuses on the perception components of the car, and therefore it does not need to include a specification of the vehicle's control, but requires a refined, more precise representation of the map.
Both models were not yet fully disclosed.

We validated both models using different approaches supported by the CADP tools.
First, we checked several safety and liveness properties characterizing the correct behavior of an AV.
Concretely, we expressed the properties in the MCL~\cite{Mateescu-Thivolle-08} data-handling, action-based temporal property language, and verified them on the models using the on-the-fly model checker of CADP.
Second, we generated several relevant AD scenarios from the models.
More precisely, we used TESTOR~\cite{Marsso-Mateescu-Serwe-18}, a tool developed on top of CADP for on-the-fly conformance test case generation guided by test purposes, to generate abstract test cases, which were automatically translated into AD scenarios~\cite{Horel-Laugier-Marsso-Mateescu-Muller-Paigwar-Renzaglia-Serwe-22}.

The rest of this paper is organized as follows. 
Section~\ref{sec:graph} presents the first formal model of an AV, including its control, and the validation of the model using safety and liveness properties.
Section~\ref{sec:grid} presents the second formal model of an AV, with a refined map, and generation of AD scenarios based on the model.
Section~\ref{sec:conclusion} compares both models and gives concluding remarks. 
Appendices~\ref{ap:graph-model} and \ref{ap:grid-model} give the complete LNT source code of the first and second model, respectively.

\section{Model focused on control}
\label{sec:graph}

\definecolor{ForestGreen}{RGB}{34,139,34}
\definecolor{bordeau}{RGB}{88,24,31}
\tikzset{
    carp/.style={draw, color=gray!96, rectangle,rounded corners,minimum width=2cm,minimum height=0.9cm,align=center,text width=2.4cm,font=\small\sffamily,fill=white!20},
    envp/.style={draw,color=ForestGreen,rectangle,rounded corners,minimum width=2cm,minimum height=0.9cm,align=center,text width=2.2cm,fill=ForestGreen!40,font=\small\sffamily},
    obstaclep/.style={draw,color=red, rectangle,rounded corners,minimum width=2cm,minimum height=0.9cm,align=center,text width=2cm,fill=red!20,font=\small\sffamily}
}

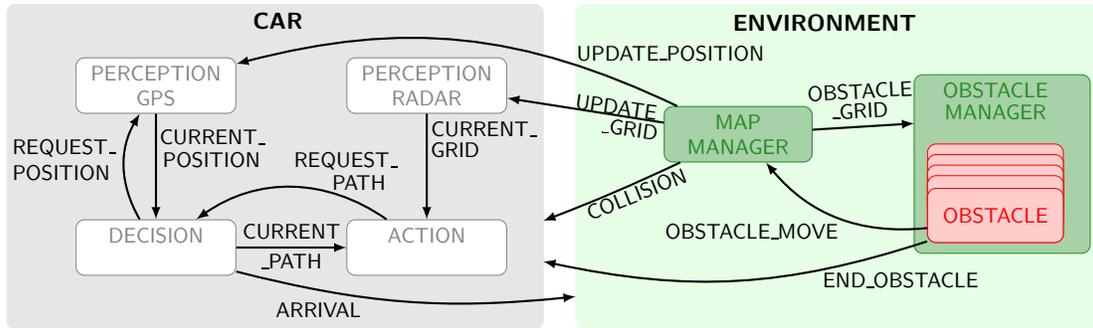
\begin{figure*}
    \centering
\scalebox{.80}{
\begin{tikzpicture}[x=2.25cm,y=0.9cm]
\draw[rounded corners, fill=gray!20, color=gray!20] (-1.2,4) rectangle (2.76,-2);
\draw[rounded corners, fill=green!10, color=green!10] (3,4) rectangle (6.9,-2);

\node[carp] (gps) at (-0.1,2.5) {PERCEPTION~\\GPS};
\node[carp, below = 50pt of gps] (decision) {DECISION\\~};
\node[carp] (radar) at (1.9,2.5) {PERCEPTION~\\RADAR};
\node[carp, below = 50pt of radar] (action) {ACTION\\~};

\node[envp] (map) at (4.2,1.6) {MAP~\\MANAGER};

\draw[rounded corners, color=ForestGreen, fill=ForestGreen!40] (5.5,2.7) rectangle (6.8,-0.6);
    \node[obstaclep] (o0) at (6.1,0.1) {~\\[1ex]~};
    \node[obstaclep, above = -5pt of o0] (j_tc3) {\\[1ex]~};
    \node[obstaclep, above = -10pt of o0] (j_tc2) {\\[1ex]~};
    \node[obstaclep, above = -15pt of o0] (j_tc1) {\\[1ex]~};
    \node[obstaclep, above = -20pt of o0] (j_tc) {\\[1ex]~};
    \node[obstaclep] (obst) at (6.1,0.1)  {OBSTACLE};

\node[rotate=0,font=\sffamily] at (0.8,3.7) {\textbf{CAR}};
\node[rotate=0,font=\small\sffamily] at (2.32,1.7) {CURRENT\_};
\node[rotate=0,font=\small\sffamily] at (2.11,1.31) {GRID};
\node[rotate=0,font=\small\sffamily] at (1.3,1.1) {REQUEST\_};
\node[rotate=0,font=\small\sffamily] at (1.4,0.7) {PATH};
\node[rotate=0,font=\small\sffamily] at (0.9,-0.2) {CURRENT};
\node[rotate=0,font=\small\sffamily] at (0.9,-0.75) {\_PATH};
\node[rotate=0,font=\small\sffamily] at (-0.78,1.35) {REQUEST\_};
\node[rotate=0,font=\small\sffamily] at (-0.8,0.9) {POSITION};
\node[rotate=0,font=\small\sffamily] at (0.315,1.55) {CURRENT\_};
\node[rotate=0,font=\small\sffamily] at (0.31,1.14) {POSITION};
\node[rotate=0,font=\small\sffamily,color=ForestGreen] at (6.1,2.41) {OBSTACLE};
\node[rotate=0,font=\small\sffamily,color=ForestGreen] at (6.1,2.01) {MANAGER};
\node[rotate=0,font=\small\sffamily] at (1.1,-1.65) {ARRIVAL};
\node[rotate=0,font=\small\sffamily] at (3.7,3.1) {UPDATE\_POSITION};
\node[rotate=-8,font=\small\sffamily] at (3.3,2.1) {UPDATE};
\node[rotate=-8,font=\small\sffamily] at (3.4,1.7) {\_GRID};
\node[rotate=3,font=\small\sffamily] at (5.1,2.4) {OBSTACLE};
\node[rotate=3,font=\small\sffamily] at (5.1,2.04) {\_GRID};
\node[rotate=0,font=\small\sffamily] at (4.3,-0.2) {OBSTACLE\_MOVE};
\node[rotate=22,font=\small\sffamily] at (3.45,0.5) {COLLISION};
\node[rotate=0,font=\small\sffamily] at (5.4,-1.1) {END\_OBSTACLE};
\node[rotate=0,font=\sffamily] at (4.84,3.67) {\textbf{ENVIRONMENT}};
\begin{scope}[every path/.style={-latex}]
\draw [line width=1pt] (gps) edge (decision)
                       (decision) edge [bend left] (gps)
                       (radar) edge (action)
                       (action) edge [bend right=35] (decision)
                       (decision) edge (action)
                       (map) edge (5.5,1.8);
\draw [line width=1pt] (map) edge (radar)
                       (map) edge [bend right=20](gps)
                       (decision) edge [bend right=10] (3, -1.4)
                       (obst) edge [bend left=28](map)
                       (map) edge [bend left=2](2.76,0)
                       (obst) edge [bend left=15](2.76,-0.75);

\end{scope}
\end{tikzpicture}
}
\vspace{-1ex}

\caption{Architecture of the autonomous vehicle LNT model focused on control}
\label{fig:graph-model}

{Arrows indicate messages sent between components and arrow labels denote the gates used.}
\end{figure*}

In the first model, the autonomous car itself consists of four components: a GPS, a radar, a decision (or trajectory) controller, and an action controller.
We chose these four components in order to represent the essential functionalities present in an autonomous car: perception (GPS and radar), decision, and action. 
The GPS keeps the current position of the car updated.
The radar detects the presence of the obstacles close to the car and builds a perception grid summarizing information about perceived obstacles.
The decision controller computes an itinerary from the current position to the destination, avoiding streets containing obstacles.
The action controller commands the engine and direction to follow the itinerary computed by the decision controller, using the perception grid built by the radar to avoid collisions.
As shown in Figure~\ref{fig:graph-model} these four components communicate in various ways: the GPS sends the current position to the decision controller upon request, the radar periodically sends the perception grid to the action controller, and the action controller requests a new itinerary from the decision controller.
This LNT model is a translation of the GRL model~\cite[Appendix~B]{Marsso-19} used to illustrate the combination of synchronous and asynchronous test generation tools~\cite{Marsso-Mateescu-Parissis-Serwe-19} to validate GALS (Globally Asynchronous, Locally Synchronous) systems.

\subsection{Elements and processes composing the model}

The car has an initial position and a destination.
Each obstacle has an initial position and a list of moves.
The model's behavior is defined such that inevitably either the car arrives, a collision occurs between the car and an obstacle, or all obstacles finish their list of moves.
The LNT model is generic and is instantiated for a particular scene by providing global constants for the map, the initial position and destination of the car, and the set of obstacles with their initial positions and lists of moves.

\paragraph{Car.}
Each component of the car is specified as a LNT process, and a process \lstinline+CAR+ defines the overall behavior as the parallel composition of these four processes (\lstinline+PERCEPTION_RADAR+, \lstinline+PERCEPTION_GPS+, \lstinline+DECISION+, and \lstinline+ACTION+) as shown in the following LNT fragment (the visible gates of a process are specified between square brackets ``\lstinline+[+...\lstinline+]+''; a process must synchronize on the gates before the arrow ``\lstinline+->+''):
\begin{lstlisting}
par
  CURRENT_GRID ->
    PERCEPTION_RADAR [UPDATE_GRID, CURRENT_GRID]
||
  REQUEST_POSITION, CURRENT_POSITION ->
    PERCEPTION_GPS [UPDATE_POSITION, REQUEST_POSITION, CURRENT_POSITION]
||
  REQUEST_PATH, CURRENT_PATH, REQUEST_POSITION, CURRENT_POSITION ->
    DECISION [REQUEST_PATH, CURRENT_PATH, REQUEST_POSITION, 
              CURRENT_POSITION, ARRIVAL] (map, destination)
||
  CURRENT_PATH, CURRENT_GRID, REQUEST_PATH ->
    ACTION [REQUEST_PATH, CURRENT_PATH, CURRENT_GRID,CAR_MOVE,COLLISION]
end par
\end{lstlisting}

\paragraph{Car components.}
The LNT process \lstinline+PERCEPTION_RADAR+ has two local variables to keep track of the current and previous perception grid.
A perception grid is represented by a list of edges corresponding to the streets occupied by the obstacles.
This grid is initially empty (i.e., it has the value \lstinline+Radar ({})+).
After each obstacle or car move, the radar receives the current grid as a message %(through the gate \lstinline+UPDATE_GRID+)
from the environment, but informs the driving controller only in case of a change% (through the gate \lstinline++CURRENT_GRID+)
.

The LNT process \lstinline+PERCEPTION_GPS+ initializes the car position, receives updates of the car position (on gate \lstinline+UPDATE_POSITION+) and position requests from the decision controller (on gate \lstinline+REQUEST_POSITION+).
Upon request, it sends the car position to the decision controller (on gate \lstinline+CURRENT_POSITION+).

The LNT process \lstinline+DECISION+ has two value parameters: the initial \lstinline+map+ and the \lstinline+destination+ of the car. Process \lstinline+DECISION+ waits for receiving from process \lstinline+ACTION+ (on gate \lstinline+REQUEST_PATH+) a request to compute a path avoiding obstacles, then requests the current position from process \lstinline+PERCEPTION_GPS+ (on gate \lstinline+REQUEST_POSITION+).
When process \lstinline+DECISION+ receives the current position (on gate \lstinline+CURRENT_POSITION+), it checks if the car arrived at the destination, in which case it performs a rendezvous on gate \lstinline+ARRIVAL+ and stops; otherwise it computes an itinerary (using classical graph exploration algorithms) and sends it to process \lstinline+ACTION+ (on gate \lstinline+CURRENT_PATH+). 
An itinerary is a list of controls, i.e., a turn in one of the crossroads (\lstinline+turned_n (N: Nat)+) or a brake (\lstinline+brakes+).
If there is no possible itinerary avoiding the obstacles, an empty itinerary is sent (if the obstacles have finished their moves, this leads to a deadlock).

The LNT process \lstinline+ACTION+ requests an itinerary avoiding the obstacles from process \lstinline+DECISION+ (on gate \lstinline+REQUEST_PATH+), and then it checks if the itinerary is feasible: if yes, it moves the car according to the first control in the itinerary (e.g., turn in the 5th crossroad), if not it waits for obstacles moves. Finally, after each perception change received from process \lstinline+PERCEPTION_RADAR+, it requests a new itinerary with the updated list of obstacle positions (on gate \lstinline+REQUEST_PATH+).

\paragraph{Geographical map.}
\begin{figure}
    \centering
    \begin{tabular}{c@{\hspace{3em}}c}
        \includegraphics[scale=0.09]{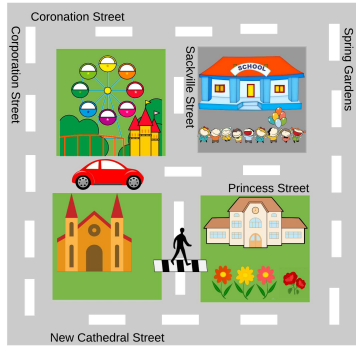} &
        \includegraphics[scale=0.6]{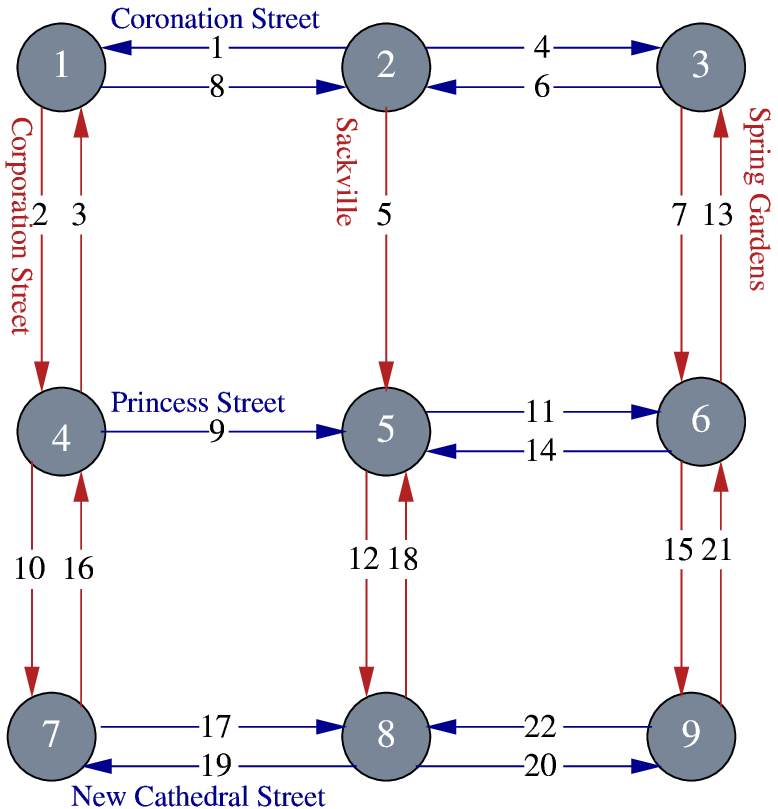} 
        \\
        (A) Map (with the car and a crossing pedestrian) &
        (B) Map representation as directed graph
    \end{tabular}
    \caption{Example of a geographical map and its corresponding graph representation}
    \label{fig:graph-map}
\end{figure}
The geographical map is represented as a directed graph as illustrated in Figure~\ref{fig:graph-map}.B, in which edges correspond to streets and nodes correspond to crossroads; for simplicity, we assume that an actor (car or an obstacle) occupies a street completely (a longer street can be represented by several edges in the graph).
A set of functions is defined to explore this graph, to compute itineraries, etc.
The example below shows a a fragment of the LNT constant~\lstinline+initial_map+ that returns the graph corresponding to the map illustrated on Figure~\ref{fig:graph-map}.
\begin{lstlisting}[language=LNT]
   function initial_map : Graph is
      var g: Graph, e: Edges, v: Vertices in
         v := {0, 1, 2, 3, 4, 5, 6, 7, 8};
         e := {Edge (0, Coronation_Street, 1), 
               Edge (0, Corporation_Street, 3), 
               Edge (1, Coronation_Street_bis, 0),
               ... };
         g := Graph (v, e);
         return g
      end var
   end function 
\end{lstlisting}

\paragraph{Obstacles.}
As the car, obstacles move from a street-segment to an (adjacent) street-segment.
If an obstacle is on the same segment as the car, this corresponds to a collision.
To limit the complexity, each obstacle executes a fixed number of random or statically chosen moves. More precisely, there are three types of obstacle moves: (1) the obstacle can either leave, (2) turn in one of the crossroads, or (3) perform a random move.
These moves are defined by the following LNT type \lstinline+Operation+.
\begin{lstlisting}[language=LNT]
   type Operation is -- obstacle operations
      random,
      turned_n (N: Nat),
      leave
   end type
\end{lstlisting}
The behaviour of an obstacle is specified in the process \lstinline+OBSTACLE+.
It consists of executing the sequence of the obstacle's moves and then stop. 
The first move corresponds to the appearance of the obstacle at its initial position.
For \lstinline+random+, the next move is chosen randomly among those possible in the current map (e.g., leave or turn (0)) as shown in the following LNT fragment of the process \lstinline+OBSTACLE+.
\begin{lstlisting}[language=LNT]
   if opi == random then
      select
         opi := leave
      [] var n0, n1: Nat in
            n1 := length (succ_l (map.G.E, o.position));
            n0 := any Nat where n0 <= n1;
            opi := turned_n (n0)
         end var
      end select
   end if;
\end{lstlisting}
Note that the LNT operator \lstinline+select+ specifies non-deterministic choice.
Finally, when the next move is chosen, the obstacle only executes the move if the destination segment is free; otherwise, it waits.
Note that the process \lstinline+OBSTACLE+ receives from the process \lstinline+ENVIRONMENT+ updates of the positions of the car and obstacles.
The LNT process \lstinline+OBSTACLES_MANAGER+ groups several obstacles in a parallel composition, providing the initial list of moves to each obstacle.

\paragraph{Map management.}
The map is akin to a global variable modified at each move of the car or an obstacle.
The updates of the map are handled by the process \lstinline+MAP_MANAGEMENT+, which ensures that:
(1) the geographical map information, such as the position of the car (\lstinline+map.c+) and obstacles (\lstinline+grid+), is shared with the process \lstinline+RADAR+ (by sending this information to \lstinline+RADAR+ on gate \lstinline+POSITIONS+);
(2) the geographical map information is updated when the car or the obstacles move (by receiving these moves from the processes \lstinline+ACTION+ and \lstinline+RADAR+)%
; and
(3) the processes can only be executed as long as the car did neither arrive at destination nor crash, and at least one obstacle has still some moves to execute.
The environment is defined as the parallel composition of the process \lstinline+OBSTACLES_MANAGER+ and \lstinline+MAP_MANAGEMENT+ in the process \lstinline+ENVIRONMENT+.

\subsection{Module hierarchy and scenario module}

The model is split in three different parts.
First, a part with definitions independent of the AV, e.g., types related to graph definitions together with the classical graph functions.
Second, a generic part, defining types, functions, and processes common to all configurations.
Finally, a particular part, defining the constants characterizing the considered configuration: the geographical map, the number of obstacles, as well as the initial position and behaviour of these obstacles.

\paragraph{Stats.}

The resulting LNT specification has 881 lines dispatched in four modules, containing 14~types, 19~functions, six channels, and ten~processes.
Using CADP, for the map with 22 streets and 8 crossroads represented in Figure~\ref{fig:graph-map} and two obstacles, each with one random move, we generated (in about a minute on a standard laptop) the corresponding LTS (Labelled Transition System): 59,781 states and 179,884 transitions (13,305 states and 28,601 transitions after strong bisimulation minimization).

\subsection{Validation by model checking}
We validated our LNT model by checking several safety and liveness properties characterizing the correct behavior of the AV. 
We expressed the properties in MCL~\cite{Mateescu-Thivolle-08}, the data-handling, action-based, branching-time temporal logic of the on-the-fly model checker of CADP.
We describe here two properties in natural language and in MCL---more were verified for the associated GRL model~\cite{Marsso-Mateescu-Parissis-Serwe-19}.

\paragraph{Property 1:}
\sloppypar
``\emph{The position of the car is correctly updated after each move of the car.}''
This safety property expresses that on each transition sequence, an update of the car position (``\lstinline+UPDATE_POSITION ?current_street+'', where \lstinline+current_street+ is the street on which the car currently is) followed by a car move (``\lstinline+CAR_MOVE ?control+'', where \lstinline+control+ is a move) cannot be followed by an update of the car position inconsistent with \lstinline+current_street+, \lstinline+control+, and the map.
This can be expressed in MCL using the necessity modality below, which forbids transition sequences containing inconsistent position updates:
\begin{lstlisting}[language=MCL]
   [ true* .
     { UPDATE_POSITION ?current_street:String } .
     (not ({ CAR_MOVE ... } or { UPDATE_POSITION ... }))* .
     { CAR_MOVE ?control:String } .
     (not ({ CAR_MOVE ... } or { UPDATE_POSITION ... }))* .
     { UPDATE_POSITION ?new_street:String where
       not (Consistent_Move (current_street, control, new_street)) }
   ] false
\end{lstlisting}
The values of the current position, the move, and the new position of the car occurring as offers for the gates \lstinline+UPDATE_POSITION+ and \lstinline+CAR_MOVE+ are captured in the variables \lstinline+current_street+, \lstinline+control+, and \lstinline+new_street+ of the corresponding action predicates (surrounded by curly braces) and reused in the \lstinline[language=MCL]+where+ clause of the last action predicate.
The Boolean function \lstinline[language=MCL]+Consistent_Move+ defines all valid combinations for \lstinline+current_street+, \lstinline+control+, and \lstinline+new_street+ allowed by the map.
    
\paragraph{Property 2:}  ``\emph{Inevitably, the system should reach a state where either the car arrived (\lstinline[language=MCL]+ARRIVED+), or a collision occurred between the car and an obstacle (\lstinline[language=MCL]+COLLISION+), or all obstacles have finished their list of moves (\lstinline[language=MCL]+END_OBSTACLE+).}''
This property can be expressed in MCL using the formula below, which forbids infinite transition sequences not containing one of the three terminal actions (\lstinline[language=MCL]+TERMINATE+):
\begin{lstlisting}[language=MCL]
   not <(not TERMINATE)>@
\end{lstlisting}
Here, \lstinline[language=MCL]+TERMINATE+ encompasses all three terminal actions \lstinline[language=MCL]+ARRIVED+, \lstinline[language=MCL]+COLLISION+, and \lstinline[language=MCL]+END_OBSTACLE+.
Note that this formula correctly expresses the property if and only if the model is free of deadlocks, which can be easily verified by another check.

Note that we also specified test purposes and generated test cases from this model using TESTOR \cite{Marsso-Mateescu-Serwe-18} as part of the evaluation of a test generation approach~\cite{Marsso-Mateescu-Serwe-20}.

\subsection{Discussion}
We specified an AV close to a deployed one, i.e., we considered an AV as consisting of several components for action, decision, and perception.
To naturally specify the search of an itinerary, we represented the geographical map as a graph, and formalized classical graph exploration algorithms.
This model is a translation of a GALS model~\cite{Marsso-Mateescu-Parissis-Serwe-19,Marsso-19} previously described in the formal GRL language~\cite{Jebali-Lang-Mateescu-16,Jebali-16}, where the various components of the car are represented as synchronous programs interacting globally in an asynchronous manner.

This first model can directly be used to test the control of an AV, since it computes the possible itineraries avoiding collisions.
However, the abstraction of each component can be refined according to different evaluation goals.
For instance, to only test the perception component (e.g., radar) the current formal model is too general and the current abstraction of the map is not optimal to verify a perception component.
In particular, more precision is required for the geographical map (e.g., concrete angles and lengths of street segments) and (the trajectories of) obstacles (e.g., their size and speed).
On the other hand, some information might not be necessary to test a particular component (e.g., the control and action processes might be irrelevant to test the perception): thus, these aspects can be dropped from the model, consequently improving validation performance by reducing the size of the underlying LTS.

In the next section, we will present a second formal model for an AV.
If the overall structure of the second model is the same (in particular the module hierarchy and the processes \lstinline+OBSTACLES_MANAGER+ and \lstinline+MAP_MANAGEMENT+), the second model focuses on the perception component, which requires a more precise representation of the geographical map.

\section{Model focused on perception}
\label{sec:grid}

To validate the perception components, which are crucial for an AV, it is necessary to devise a model focused on perception aspects. We present below such a model in LNT, derived from the control-focused model given in Section~\ref{sec:graph}. This second model keeps the same management of actors and important events (arrival of the car at destination, end of obstacle trajectories, and collision), but represents the map using an array instead of a graph, and abstracts away most car components, except for the perception LiDAR. This perception-focused model was used to generate scenarios to be executed on an AD simulator.

\subsection{Elements and processes composing the model}

\definecolor{ForestGreen}{RGB}{34,139,34}
\definecolor{bordeau}{RGB}{88,24,31}
\tikzset{
    carp/.style={draw, color=gray!96, rectangle,rounded corners,minimum width=2cm,minimum height=0.9cm,align=center,text width=2.4cm,font=\small\sffamily,fill=white!20},
    envp/.style={draw,color=ForestGreen,rectangle,rounded corners,minimum width=2cm,minimum height=0.9cm,align=center,text width=2.2cm,fill=ForestGreen!40,font=\small\sffamily},
    obstaclep/.style={draw,color=red, rectangle,rounded corners,minimum width=2cm,minimum height=0.9cm,align=center,text width=2cm,fill=red!20,font=\small\sffamily},
    schp/.style={draw,color=blue!90,rectangle,rounded corners,minimum width=2cm,minimum height=0.9cm,align=center,text width=2.2cm,fill=blue!20,font=\small\sffamily},
    cir/.style={draw,color=black,circle,fill=black,font=\small\sffamily}
}

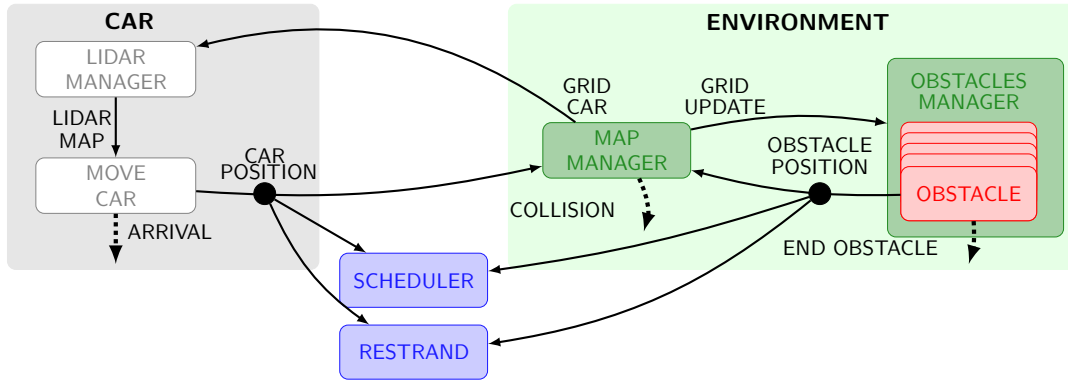
\begin{figure*}
    \centering
\scalebox{.80}{
\begin{tikzpicture}[x=2.25cm,y=0.9cm]
\draw[rounded corners, fill=gray!20, color=gray!20] (-1,4) rectangle (1.3,-0.9);
\draw[rounded corners, fill=green!10, color=green!10] (2.7,4) rectangle (6.9,-0.9);

\node[carp] (lidar) at (-0.2,2.8) {LIDAR~\\MANAGER};
\node[carp, below = of lidar] (move) {MOVE\\CAR};

\node[schp] (scheduler) at (2,-1.1) {SCHEDULER};
\node[schp,below = 8pt of scheduler] (restrand) {RESTRAND};

\node[cir] (br1) at (0.9,0.5) {~};
\node[cir] (br2) at (5,0.5) {~};

\node[envp] (map) at (3.5,1.3) {MAP~\\MANAGER};

\draw[rounded corners, color=ForestGreen, fill=ForestGreen!40] (5.5,3) rectangle (6.8,-0.3);
\node[obstaclep] (o0) at (6.1,0.5) {~\\[1ex]~};
    \node[obstaclep, above = -5pt of o0] (j_tc3) {\\[1ex]~};
    \node[obstaclep, above = -10pt of o0] (j_tc1) {\\[1ex]~};
    \node[obstaclep, above = -15pt of o0] (j_tc2) {\\[1ex]~};
    \node[obstaclep, above = -20pt of o0] (j_tc) {\\[1ex]~};
    \node[obstaclep] (obst)  at (6.1,0.5)  {OBSTACLE};

\node[rotate=0,font=\sffamily] at (-0.1,3.7) {\textbf{CAR}};
\node[rotate=0,font=\small\sffamily,color=ForestGreen] at (6.1,2.59) {OBSTACLES};
\node[rotate=0,font=\small\sffamily,color=ForestGreen] at (6.1,2.2) {MANAGER};
\node[rotate=0,font=\small\sffamily] at (4.3,2.5) {GRID};
\node[rotate=0,font=\small\sffamily] at (4.3,2.1) {UPDATE};
\node[rotate=0,font=\small\sffamily] at (3.28,2.5) {GRID};
\node[rotate=0,font=\small\sffamily] at (3.28,2.1) {CAR};
\node[rotate=0,font=\small\sffamily] at (5,1.4) {OBSTACLE};
\node[rotate=0,font=\small\sffamily] at (5,1) {POSITION};
\node[rotate=0,font=\small\sffamily] at (3.1,0.2) {COLLISION};
\node[rotate=0,font=\small\sffamily] at (0.2,-0.2) {ARRIVAL};
\node[rotate=0,font=\small\sffamily] at (0.9,1.2) {CAR};
\node[rotate=0,font=\small\sffamily] at (0.93,0.9) {POSITION};
\node[rotate=0,font=\small\sffamily] at (-0.45,1.9) {LIDAR};
\node[rotate=0,font=\small\sffamily] at (-0.45,1.48) {MAP};
\node[rotate=0,font=\small\sffamily] at (5.3,-0.5) {END OBSTACLE};
\node[rotate=0,font=\sffamily] at (4.84,3.67) {\textbf{ENVIRONMENT}};
\begin{scope}[every path/.style={-latex}]
\draw [line width=1pt] (lidar) edge (move)
                       (move) edge [bend right=8](map)
                       (map) edge [bend right=25] (lidar)
                       (map) edge [bend left=10](5.5,1.8)
                       (obst) edge [bend left=8](map)
                       %(map) edge (3.5,-0.3)
                       %(obst) edge (6.1,-0.7)
                       %(move) edge (-0.2,-0.38)
                       (br1) edge (scheduler)
                       (br1) edge [bend right =15] (restrand)
                       (br2) edge [bend left=5](scheduler)
                       (br2) edge [bend left =15] (restrand);

\draw[dotted,line width=2pt]
      (map) edge [bend left=20](3.7,-0.2)
      (obst) edge [bend left=10](6.1,-0.8)
      (move) edge (-0.2,-0.8);

\end{scope}
\end{tikzpicture}
}
\vspace{-1ex}

\caption{Architecture of the autonomous vehicle LNT model focused on perception}
\label{fig:processes}
\end{figure*}

    \paragraph{Constants and Channels.}
    The principal constants are the height and width of the array defining the map, and those of the array defining the perception grid of the \lstinline+LIDAR+ component. They are mostly used to verify if an actor is present on the map or not. The default size of the map is ten$\times$ten cells, and it can be increased for a higher resolution. The various processes in the model communicate through gates defined with channels used to convey specific data values. We defined seven different channels in the model.

     %-- Obstacles : can be one cell or more, different constant speed, have a set of movement in a trajectory, can choose to take a movement or not, can have the choice of the movement into the automaton, is transparent or not, pedestrian or other car. Can have diagonal movement but it does not depend on the speed of the obstacle. Mention the name of the obstacles that are unique and are done with the type avoiding string use. Random direction a restrained because of the need to have less state\\
     %-- channel O is (Obstacle, Obstacle, Direction) end channel -- Obstacle, Obstacle, Dir\\
     %-- Processes updating obstacles' positions and behaviour depending on the grid and the position of the other actor. Also depending on the next movement given by the trajectory.\\
     \paragraph{Obstacles.}
     %Type definition obstacles
     The obstacles are actors on the map that represent various objects and elements of the environment. Obstacles can have various sizes (minimum one cell) as they can be pedestrians, cars, or buildings. Each obstacle has a unique identifier defined as a value of an enumerated type. Obstacles can be static or mobile, the latter ones being able to move in any direction. Each move consists in traversing a number of cells determined by the obstacle speed. Some obstacles can hide the view (e.g., a car or a building) and some cannot (e.g., a pedestrian or a pole). 
     
     In a configuration, each obstacle has a list of moves defining its behavior. Its moves depend on its speed and direction. An obstacle may also choose not to move or may randomly choose between several possible directions, which adds nondeterminism to the model and enables the exploration of further scenarios. As in the first model, obstacle moves must not lead to collisions (i.e., end on occupied cells of the map) in order to yield relevant scenarios. To keep the model size tractable, we enable full random moves only for obstacles close enough to the car, the random moves of the farther obstacles being restricted to directions bringing them closer to the car.
     
     Each obstacle is modeled by an instance of the \lstinline+OBSTACLE+ process, which has as parameters the obstacle's identifier, its list of moves, and a flag indicating whether the obstacle has a cyclic movement. Before attempting the next obstacle move (at the head of the list), process \lstinline+OBSTACLE+ obtains, via gate \lstinline+GRID_UPDATE+, the current map from the \lstinline+MAP_MANAGER+ process. Based on the map, on the current obstacle information (position and speed), and on the direction of the next move, the obstacle determines whether the move is valid, i.e., does not lead to a collision. Then, process \lstinline+OBSTACLE+ sends on gate \lstinline+OBSTACLE_POSITION+ the previous position, next position, and new direction of the obstacle (including the case when the obstacle does not move) to process \lstinline+MAP_MANAGER+, in charge of updating the map.
     If the next direction of move is random, the possible moves of the obstacle are constrained by the \lstinline+RESTRAND+ process, described later in this section. When the list of moves is finished, process \lstinline+OBSTACLE+ performs an \lstinline+END_OBSTACLE+ action and stops moving, except when it has a cyclic behaviour, in which case it starts again using its list of moves given initially. The process \lstinline+OBSTACLES_MANAGER+ is the parallel composition of all \lstinline+OBSTACLE+ processes in the considered configuration.
     
     %In case it is random, the actual direction is chosen by taking into account the obstacle situation (previous position and speed) and the state of the map.
     %If it is random, the direction chosen depends on the restraint of the obstacle. It uses a function modifying the position of the obstacle depending on its previous position, the direction, its speed and the state of the Map and send the updated position through the gate \lstinline+OBSTACLE_POSITION+.

\begin{figure}
    \centering
    \includegraphics[scale=0.5]{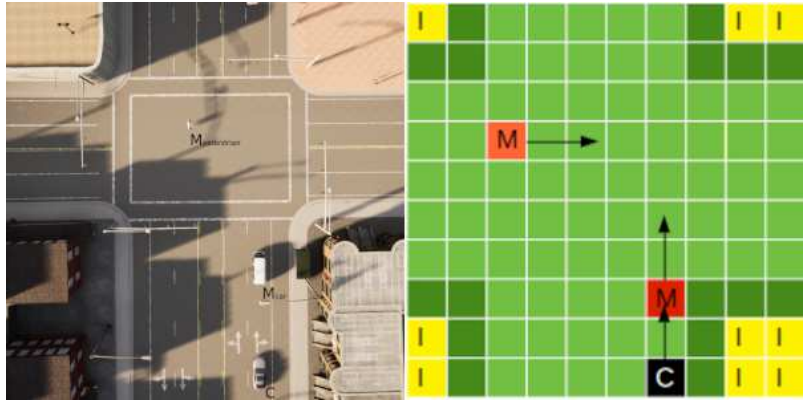}
    \caption{Representation of the map with three actors (car and two mobile obstacles)}
    \label{fig:map_grid}
\end{figure}
     %-- Grid, variable size, set of cells which can have different value ( free, occupied by and obstacle, occupied by the car), the   grid can be variable to set different type of environment. Set with first position of each obstacles.\\
    \paragraph{Map.}
    %Code type grille et function initialisation
    The map is essentially the representation of the ground truth, the world in which the different actors move. The map is represented as a 2--dimensional array composed of cells with different values: \lstinline+free+ when there is no obstacle nor car on the cell, \lstinline+occupied (obstacle)+ when the cell is occupied by an obstacle (including all the obstacle information), and \lstinline+car_pos+ when the cell is occupied by the car (we consider only one car on the map). The map is defined and initialized in LNT as follows:
    \begin{lstlisting}[language=LNT]
type Map_Info is free, car_pos, occupied (O:Obstacle) end type
type Map_Line is array [0..9] of Map_Info end type
type Map is array [0..9] of Map_Line end type

function initiate_map (out var m:Map) is
  -- create an empty map
  m := Map (Map_line (free));
  -- add static obstacles into the ground truth array
  add_obstacle (!?m, Obstacle (Scenery,
                               CellRectPosition (Position (0, 0)),
                               Obstacle_Speed (0), none, false));
  ...
  -- add mobile obstacles into the ground truth array
  add_obstacle (!?m, Obstacle (Other_Car,
                               CellRectPosition (Position (6, 7)),
                               Obstacle_Speed (1), up, false));
  ...
end function
\end{lstlisting}
      
    % Radu: pas absolument necessaire de mentionner cela, on se doute bien qu'il y a de bonnes raisons de separer le car et les obstacles 
    % We differentiate the car and the obstacles because it is easier to read in the code and the information needed are different. The car is only represented by its position, and the obstacles have different attributes. 
    \noindent
    The \lstinline+MAP_MANAGER+ process is a central part of the model, in charge of maintaining the map (ground truth) and of communicating with the other processes to update the position of the actors. The map is initialized with the positions of static obstacles and the initial positions of mobile obstacles. It is sent on gate \lstinline+GRID_UPDATE+ to the \lstinline+OBSTACLE+ processes to determine their next moves, and is also sent on gate \lstinline+GRID_CAR+ (accompanied by the car position) to the process \lstinline+LIDAR_MANAGER+ to generate the perception grid. If at some moment the position of the car becomes the same as one of the obstacles, \lstinline+MAP_MANAGER+ performs an action \lstinline+COLLISION+ and stops, entailing the termination of the whole scenario. The \lstinline+ENVIRONMENT+ process is the parallel composition of \lstinline+MAP_MANAGER+ and \lstinline+OBSTACLES_MANAGER+.

     %-- Process that update the grid depending on the different position the other processes send. Each process has a local value of the grid sent by this process so it can update the value accordingly but only the map\_manager can update the grid\\

    %-- Car : has a defined trajectory, a constant speed. Can make it choose its trajectory depending on purpose (for decision like test), can change its speed, occupy only one cell for the moment, scenario ends if car ends its movement or collide with an obstacle\\
    %-- compared to the obstacle the car can lead to a collision as it can go wherever it wants 
    %-- channel C is (Position, Position) end channel used to send the position of the Car and its previous position, so the previous position can be erased and the current can be added in the grid.\\
    %-- Processes updating car' positions and behaviour depending on the grid and the position of the other actor. Also depending on the next movement given by the trajectory.\\
    \paragraph{Car.}
    The \lstinline+CAR+ process is the parallel composition of the \lstinline+LIDAR_MANAGER+ and \lstinline+CAR_MOVE+ processes, the latter being in charge of managing the car moves. Process \lstinline+CAR_MOVE+ has as parameters the car's information (position, list of moves, speed, and a flag indicating if the car has a cyclic movement) and is similar to process \lstinline+OBSTACLE+. The car moves essentially in the same way as the obstacles, except that it can also move to an occupied cell, and thus trigger a \lstinline+COLLISION+ action from the \lstinline+GRID_MANAGER+ process. Upon each move of the car, its previous and current positions are transferred to the \lstinline+MAP_MANAGER+ process on gate \lstinline+CAR_POSITION+ to be updated on the map. If the car has finished its list of moves, it performs an action \lstinline+ARRIVAL+, which terminates the scenario.
    
    % Radu: pas essentiel
    % The car is represented as a value of type \lstinline+position+, which indicates the two coordinates of the cell occupied by the car. 
    
    % Radu: utile la` ou` on parle du module scenario
    % The information about the car is given in a scenario module as an argument to process \lstinline+MOVE_CAR+. 

     %--  : and grid 5x5 around the car, the center cell is always the car position, change depending on the current grid and current position of car, hidden cells behind not transparent obstacles.\\
     %channel L is (P, Grid\_Map) end channel -- LiDAR
    %-- Process LiDAR description\\
    \paragraph{LiDAR.}
    The perception grid represents the perception of the car (as computed by the LiDAR) up to a certain distance. It is modeled as a 2-dimensional array centered on the car position and having a default size of 5x5 cells. The cells of the perception grid have different values from those of the map: \lstinline+F+ for free cells, \lstinline+C+ for the car position, \lstinline+O+ for occupied cells, \lstinline+M+ for cells that were free on the last grid but became occupied, \lstinline+T+ for cells occupied by a transparent obstacle, \lstinline+N+ similar to \lstinline+M+ but for cells occupied by transparent obstacles, and \lstinline+U+ for unknown cells, i.e., those out of the map (if the grid exceeds the map boundaries) or those hidden from view (behind an opaque obstacle). The perception grid is defined as follows and initialized as an empty array: 

    \begin{lstlisting}[language=LNT]
    type Perception_Info is C, F, O, M, T, N, U end type
    type L is array [0..4] of Perception_Info end type
    type P is array [0..4] of L end type
    \end{lstlisting}
    
    \noindent
    There are several functions calculating the perception grid. The principal one takes the map, the previous perception grid and the current position of the car to calculate the new grid by translating the values of map cells into the values of the corresponding grid cells. Other functions calculate the hidden cells depending on the presence of opaque obstacles around the car, and update the value of grid cells from \lstinline+F+ to \lstinline+M+ or to \lstinline+N+ if these changed between two consecutive grids.
    % Radu: details pas absolument necessaires
    % For now, each cell is taken separately to calculate hidden cells. For a 5x5 grid, it is useless to take into account the cells on the border not hiding the rest. When we will allow a bigger size of the grid, the function will be optimized. 
    % Another function is used to compare the newly calculated grid to the previous one and change the value of cells from O to M and from T to N if one of those cells has a different value in the two configurations.
    The perception grid is maintained by the \lstinline+LIDAR_MANAGER+ process, which sends on gate \lstinline+LIDAR_MAP+ the new value of the grid and map to process \lstinline+MOVE_CAR+ to compute the next car moves. Process \lstinline+LIDAR_MANAGER+ also has a special parameter to indicate whether the contents of the grid must be kept available in the LTS for validation purposes.
    % that can be changed if we want to have an observation of the grid in the model.
    
    % Radu: pas absolument besoin de ces details
    % The process \lstinline+LIDAR_MANAGER+ creates an empty \lstinline+grid+ acting as the previous value for the first time, enters a loop with the first action updating the previous grid, calculates the new value of the grid and send the new value. 
    
    \paragraph{Scheduler and Restrand.} 
    Two auxiliary processes optimize the model regarding both its scalability and its realism when connected to an AD simulator. The \lstinline+SCHEDULER+ process introduces additional synchrony in the model to bring it closer to its physical counterpart, by enforcing that between two consecutive time instants, every actor must perform one move (at its own speed). Process \lstinline+SCHEDULER+ monitors the moves of the car and the obstacles by synchronizing on gates \lstinline+CAR_POSITION+ and \lstinline+OBSTACLE_POSITION+, indicates consecutive time periods by emitting signals on a special gate \lstinline+TICK+, and forces all actors (in a fixed order) to perform one move between two \lstinline+TICK+ actions (note that an obstacle may choose not to move, which amounts to keep its position unchanged). This improves the execution of transition sequences in the AD simulator, by allowing all actor moves between two \lstinline+TICK+ actions to be performed in parallel, yielding realistic movements, as opposed to jerky ones induced by equivalent, but less realistic interleavings in the absence of \lstinline+TICK+ actions. This also reduces the size of the LTS by pruning redundant interleavings of actions (equivalent orderings of obstacle moves between two \lstinline+TICK+ actions). Between two \lstinline+TICK+ actions, the car moves after all obstacles, because at the end of the scenario (car arrival or a collision), the obstacles must still be able to finish their moves in the simulator.

    %Processus example code
    % Those are two processes created to optimize the model (i.e., reduce the state space). They are not used to model one the the element but to arrange some of their behaviour. The first one is a process \lstinline+SCHEDULER+, limiting the movement of each actor and allowing synchronous composition. It is specified in LNT as follows:
    
\begin{lstlisting}[language=LNT]
process SCHEDULER [OBSTACLE_POSITION:O, CAR_POSITION:C, TICK:none] is
  loop
    OBSTACLE_POSITION (?Obstacle (Pedestrian, any RectPosition,
                       any Obstacle_Speed, any Direction, true),
                       ?any Obstacle, ?any Direction);
    ... -- all the other obstacles
    CAR_POSITION (?any Position, ?any Position);
    TICK
  end loop
end process
\end{lstlisting}
    
    \noindent
    % Between each \lstinline+TICK+, every actor moves once. It is defined in the simulator as a simultaneous move, as if it was paralleled. It avoids sequential movement and keep the realism of the result. It also limits the number of states of the resulting automaton when an obstacle has an infinite movement (and then can move an infinity of time between two car movement). The car action is done at the end because if the scenario ends, the other obstacles must be able to finish their movement as well in the simulator.
    %The CAR movement was put before the obstacle movement, but it led to a problem when it was converted to CARLA, as the scenario ends when the car collide with an obstacle, theses obstacles does not have the time to finish their movement in the model, and then they will not move in the last "turn" where all the obstacles are to move at the same time. 
    
    The \lstinline+RESTRAND+ (for ``restraining random'') process limits the random moves of the obstacles to keep them in a meaningful neighbourhood of the car. This is useful both for specifying scenarios with relevant obstacle trajectories (obstacles close enough to be perceived by the LiDAR) and for reducing the LTS. Process \lstinline+RESTRAND+ monitors the obstacle moves by synchronizing on gate \lstinline+OBSTACLE_POSITION+ and constrains the random moves (depending on the car position, the previous obstacle position, a direction and the minimal distance) to force obstacles to get close enough to the car to be perceived.

    % The second process is a process named \lstinline+RESTRAND+ (for restraining random) which limits the movement of random obstacles as previously explained in the obstacle part. This optimisation is necessary to specify scenarios with relevant obstacle trajectories (the obstacle will get close enough for an observation) and to reduce the state space of the generated automaton. It restrains the movement of obstacles too far, forcing them to only move closer to the car. The area around the car in which an obstacle isn't restrain can be modified. The process works as a multi-way rendezvous: it uses the same gate \lstinline+OBSTACLE_POSITION+ and adds a condition on the direction value, using the keyword \lstinline+where+ and a function depending on the position of the car, the previous position of the obstacle, a direction and the minimal distance and returning true if the direction is allowed for the obstacle. This way, the transition with the wrong direction are forbidden, thus reducing the number of possible transitions.
\begin{lstlisting}
     OBSTACLE_POSITION (?any Obstacle, ?any Obstacle, dir)
        where dir != random
  [] OBSTACLE_POSITION (?obst, ?obst_prev, random)
        where moveAllowed (car, obst_prev, obst.dir, distMin)
\end{lstlisting}
    
    %The second process is a process named \lstinline+RESTRAND+ (for restraining random) limiting the movement of obstacles which can choose their movement as explained in the part about obstacles. It works that way : first we decide the area around the car in which an obstacle is allowed to move randomly and we define the initial position of the car. It uses the same gate \lstinline+OBSTACLE_POSITION+. It doesn't restrain the obstacle with a direction value other than random. It executes a function depending on the position of the car, the previous position of the obstacle, a direction and the minimal distance defined previously and return true if the direction is allowed for the obstacle. 
    %It works with a rendez vous between the value received by the MAP MANAGER process and the value received by this process, limited with the keyword \lstinline+where+ applying a condition on the value, condition given by the previously described function.

    %\vskip 0.05in
    %\noindent
    %{\bf Stats.}The resulting LNT implementation is composed of 8 modules, containing 13 types, 7 channels, 10 processes and 22 functions in 1179 lines (without taking into account the lines of the module scenario that can vary. The example in the appendix has 107 lines). Using CADP for the generation of the model for the scenario given in the example containing 2 obstacles, we generated an LTS with 4860 states, 9403 transitions in 24 seconds. The generation time and the size of the LTS can vary depending on multiple criteria (like the number of obstacle and the number of movement in their behaviour) \\
    %TO MOVE INTO AN ARRAY WITH THE STATS OF THE GRAPH BASED MODEL
    
\subsection{Module hierarchy and scenario module} 

    % In this section we discuss the hierarchy of the various LNT modules composing the model and why we organized it this way. 
    The LNT model is structured as a hierarchy of modules, depending on the type of data the functions or processes work on. All type definitions are grouped in the top module of the hierarchy, which is then separated in two branches, one for the modules concerning the environment (obstacles, map) and the other for the modules concerning the car. The main module inherits from both branches.  

    The modules are the following: \lstinline+types+ contains the data types, constants, and generic functions; \lstinline+lidar+ contains the functions and the process for the perception grid; \lstinline+car+ defines the behaviour of the car; \lstinline+map+ contains the functions for updating the map; \lstinline+obstacles+ contains the functions and processes to move obstacles; \lstinline+scenario+ contains the definitions specific to a configuration; \lstinline+map_manager+ contains the process monitoring the map; finally, the main module contains the parallel composition of the principal processes.

    % Radu: un peu trop de details ici
    % There is an additional scenario module, which contains the processes used to modify the parameters of the model. It is situated after the two branches and separates the grid module in two modules: one containing the function used to initiate and update the grid and the other containing the process \lstinline+MAP_MANAGER+ and the process \lstinline+ENVIRONMENT+, needing the initialized function situated in the scenario module while this module also needed the update functions. This way we can create different versions of the module while modifying only one file instead of modifying every file or add multiple arguments to the model.

    To easily build various configurations of the LNT model, the scenario module enables to choose the map, the initial positions of (static and dynamic) obstacles, and the behaviour of the car.
    % It is composed of one function initializing the grid and multiple processes described before, leading one to create its own scenario.
    The only parameter not defined in the scenario module is the size of the map, defined in the \lstinline+types+ module because of the hierarchy. The example \lstinline+SCENARIO+ given in Appendix~\ref{ap:grid-model} contains four static obstacles and two moving obstacles (another car and a pedestrian) with a simple trajectory. We decided to place the processes \lstinline+RESTRAND+ and \lstinline+SCHEDULER+ in the scenario module because these processes depend on the number of obstacles and, for \lstinline+RESTRAND+, on the distance from the car that is chosen for randomly moving obstacles. 
    
\paragraph{Stats.}

The resulting LNT specification has 1059 lines (excluding the scenario module, the size of which depends on the configuration---the one given in Appendix~\ref{ap:grid-model} has 161 lines) dispatched in eight modules, containing 13~types, 38~functions, seven channels, and eleven processes.
Using CADP, for the map of size ten$\times$ten represented in Figure~\ref{fig:map_grid} and two obstacles, we generated (in less than a minute on a standard laptop) the corresponding LTS with 27,168 states and 50,719 transitions (14,595 states and 28,287 transitions after strong bisimulation minimization).

  %  -- Scenario module in LNT where we chose the map, the initial position of obstacles, their number, their trajectory, the trajectory of the car, its behaviour. It helps putting every change in one module (two modules if you take into account that there is a change in the Type one about the map size and the obstacles name)\\

\subsection{Test case generation for simulation scenarios}

    The model focused on perception was devised for the synthesis of AD simulation scenarios from conformance test cases generated using the TESTOR~\cite{Marsso-Mateescu-Serwe-18} tool, as reported in~\cite{Horel-Laugier-Marsso-Mateescu-Muller-Paigwar-Renzaglia-Serwe-22}. The model was first validated by checking several temporal logic properties expressed in MCL~\cite{Mateescu-Thivolle-08}.
    % Radu: pas absolument necessaire, on se doute bien de cela
    % We then generate an LTS whose complexity depends on the number of obstacle, the number of movement and the number of movement of the car and containing every movement the actors can do as well as the corresponding configuration of the map and perception grid.
    
    The generation of test cases is guided by test purposes, which are high-level descriptions of the sequences of actions to be reached by testing. A test purpose is an LNT process specifying a sequence of actions terminated by a \lstinline+TESTOR_ACCEPT+ action characterizing the desired situation. An example of test purpose defining a sequence of (zero or more) actions leading to a \lstinline+COLLISION+ is given below.
    % Radu: pas absolument necessaire
    % Those actions do not necessarily needs to follow one another but it needs to be done in the give order.
    % We can also add a \lstinline+TESTOR_REFUSE+ transition that can stop an event from happening.

    \begin{lstlisting}[language=LNT]
    process PURPOSE [TESTOR_ACCEPT: none, COLLISION: COLLISION_O] is
      COLLISION (Pedestrian);
      loop TESTOR_ACCEPT end loop
    end process
    \end{lstlisting}

    % Radu: trop detaille
    % We then generate an LTS from a combination of the test purpose and a model named CTG (Complete Test Graph) which is an LTS containing every sequences corresponding to what was defined in the test purpose. These sequences are extracted as sequential test cases, sequential list of actions leading to the events described in the test purpose.
    % For example, if we had a test purpose with \lstinline+COLLISION(pedestrian)+ in it, every test cases generated ends with a collision with a pedestrian but with various behaviour from the obstacles and various observations.
    % The test cases generated are then translated into json configuration file usable by the simulator. 

    \noindent
    We tested the model with ten different configurations, on three different maps, with three different test purposes. The first configuration was the one represented on Figure~\ref{fig:map_grid}. On a crossroad, a car (moving obstacle) is going straight to the north, the car is following it, and a pedestrian is trying to cross and walk between them. The two different outcomes defined in test purposes are that a collision happens or not. Three other configurations were variants of this one on the same map. In two other configurations, the obstacles do not have defined trajectories, their moves being specified in the test purpose (reaching a collision or a particular perception grid). Another configuration involved a map representing a highway where three vehicles (including the car) move at different speeds in the same direction and try to change the lane. A further configuration involved a T-shaped crossroad, where another vehicle ignores the signalisation, leading to a collision with the car. Finally, we specified two configurations containing an additional obstacle that may produce near misses, where the car would be just next to an obstacle but not collide with it.
    % Radu: pas absolument necessaire
    % We deliberately kept it on reduced map size for the experiments. We also experimented on how the number of obstacles or the size of the map can change the complexity and the generation time of the LTS.

\subsection{Discussion}
This second model focuses on a particular component (i.e., the perception), with a more precise representation of the geographical map.
The advantage of this focus is the possibility to refine the precision of the moves of the obstacles and the car (e.g., by increasing the resolution of the map and perception grid) and to fine-tune the model to cover a large number of relevant AD perception scenarios. For instance, this model enables random trajectories for the obstacles with different speeds around the car, within an area of parameterized size managed by the \lstinline+RESTRAND+ process.
Although the second model focuses on the perception component, (re)integrating a control component could enrich the car's trajectory, which also impacts the perception in AD scenarios.
However, a simple control component computing a random trajectory (or executing a precomputed one) for the car would be sufficient.

\section{Conclusion}
\label{sec:conclusion}

Existing work in formal models of autonomous driving (AD) systems focuses either on modelling one component in particular as in~\cite{Ingrand-19}, or on verifying a given AD scenario as in~\cite{chen2021formal}, where a DSL is proposed to specify AD scenarios, which are subsequently translated into networks of stochastic hybrid automata and are verified using the UPPAAL-SMC model checker~\cite{Bulychev-et-al-12}.

In contrast, we presented two formal LNT models specifying the interaction of an AV with its environment, for the purpose of generating AD scenarios and testing the vehicle or some of its components in an deployment environment.
Our first model is more general, i.e., consists of a higher abstraction of both the AV and the environment, and includes several components (perception, decision, action) of the AV, whereas our second model focuses more on a particular component of the vehicle (i.e., the perception component) and considers a fragment of the geographical map, which results in a more refined representation of both the component and the geographical map representation.
Both models can be and have been used to test AVs: the first one to generate tests for the entire system, which does not require a refined abstraction of each AV component, and the second one to generate tests for validating a particular perception component, which requires a more precise representation of the map.

As future work, we plan to unify both models.
This could be beneficial for the generation of more refined test cases, i.e., including the control of the car on a precise fragment of the geographical map. 
We then plan to transform these refined test cases into scenarios for an AD simulator such as CARLA, using the approach of~\cite{Horel-Laugier-Marsso-Mateescu-Muller-Paigwar-Renzaglia-Serwe-22}.
Finally, we plan to take advantage of both the refined test cases and the corresponding derived AD scenario to test the control component using online conformance testing techniques.

~

\begin{small}
\noindent
\textbf{Acknowledgments.} This work was partly supported by project ArchitectECA2030 that has been accepted for funding within the Electronic Components and Systems for European Leadership Joint Undertaking in collaboration with the European Union's H2020 Framework Programme (H2020/2014-2020) and National Authorities, under grant agreement No.~877539.
\end{small}

\typeout{}

\clearpage
\appendix
\section{Full model focused on control}
\label{ap:graph-model}

This appendix presents the first LNT model of the autonomous vehicle (AV) and its interaction with the environment. 
This model is fully self-contained and does not depend on any externally-defined library.
For readability, the specification has been split into 8 parts, each part being devoted to a particular module, or a collection of test purpose. 
The first parts contain general definitions that could be independent of the autonomous vehicle; starting from Section~\ref{section:av-related}, the definitions become increasingly more AV-specific.

\subsection{Definitions of generic graph functions and datatypes}
\label{section:graph_car}

The module \lstinline+graph_car+ defines the types related to graph definitions together with usual  graph-exploration functions.
This module is independent of the autonomous vehicle: for a use in another context, one only needs to replace the type \lstinline+Street+ by an appropriate type characterising edge labels, e.g., the generic type \lstinline+String+ or another type specific to a domain.
The module clause \lstinline+with+ request the automatic definition of predefined functions for all types defined in the module.

\begin{lstlisting}
module graph_car
with ==, !=, get, set, inter, union
is

------------------------------------------------------------------------
--  Types
------------------------------------------------------------------------

type Street is
  Coronation_Street, 
  Corporation_Street, 
  Coronation_Street_bis,
  two_Coronation_Street,
  Sackville,
  two_Coronation_Street_bis,
  Spring_Gardens,
  Corporation_Street_bis,
  Princess_Street,
  two_Corporation_Street,
  two_Princess_Street,
  two_Sackville,
  Spring_Gardens_bis,
  two_Princess_Street_bis,
  two_Spring_Gardens,
  two_Corporation_Street_bis,
  New_Cathedral_Street,
  two_Sackville_bis,
  New_Cathedral_Street_bis,
  two_New_Cathedral_Street,
  two_Spring_Gardens_bis,
  two_New_Cathedral_Street_bis,
  -- obstacle labels:
  Lily,
  Theo,
  HIDDEN
end type

------------------------------------------------------------------------

type Corner is
  0,
  1,
  2,
  3,
  4,
  5,
  6,
  7,
  8
end type

------------------------------------------------------------------------

type Edge is
  Edge (q1: Corner, l: Street, q2: Corner)  
end type

------------------------------------------------------------------------

type Edges is
  list of Edge
with head, tail, reverse, length, member 
end type

------------------------------------------------------------------------

type Vertices is
  list of Corner
with head, tail, member
end type

------------------------------------------------------------------------

type Graph is
  -- Digraph (directed graph)
  Graph (V: Vertices, E: Edges)
end type

------------------------------------------------------------------------
--  Functions
------------------------------------------------------------------------

function add_Edge (q1, q2: Corner, l: Street, e: Edges) : Edges is
  -- add (q1, l, q2) to e
  return cons (Edge (q1, l, q2), e)
end function

------------------------------------------------------------------------

function succ_s (in var E: Edges, s: Corner) : Vertices is
  -- return targets of all edges in E leaving s 
  var list_succ: Vertices in
    list_succ := {};
    loop
      case E
      var q1, q2: Corner in
        NIL ->
          return list_succ
      | CONS (Edge (q1, any Street, q2), E) ->
          if q1 == s then 
            list_succ := cons (q2, list_succ)
          end if 
      end case
    end loop
  end var
end function

------------------------------------------------------------------------

function succ_l (in var L: Edges, e: Edge) : Edges is
  -- return all edges of E that leave the target of e 
  var list_succ: Edges in
    list_succ := {};
    loop
      case L
      var q1, q2: Corner, s: Street in
        NIL ->
          return reverse (list_succ)
      | CONS (Edge (q1, s, q2), L) ->
          if q1 == e.q2 then 
            list_succ := add_Edge (q1, q2, s, list_succ)
          end if 
      end case
    end loop
  end var
end function

------------------------------------------------------------------------

function succ_l_s (in var E: Edges, l: Street) : Corner is
  -- return target state of a succesor edge for a label
  return edge_l (E, l).q2
end function

------------------------------------------------------------------------

function edge_l (in var E: Edges, l: Street) : Edge is
  -- return edge in E with label l
  loop
    assert E != {};
    if head (E).l == l then
      return head (E)
    end if;
    E := tail (E)
  end loop
end function

------------------------------------------------------------------------

function is_l_in_E (in var E: Edges, l: Street) : Bool is
  -- return true iff there is an edge in E with label l
  loop
    case E
    var l1: Street in
      NIL ->
        return false
    | CONS (Edge (any Corner, l1, any Corner), E) ->
      if l1 == l then
        return true
      end if
    end case
  end loop
end function

------------------------------------------------------------------------

function pred_edge (E: Edges, in var D: Edge) : Edges is
  -- return all edges of E that have the source of D as target
  var pred, temp: Edges in
    pred := {D};
    temp := E;
    loop L in
      if temp == {} then
        return pred
      end if;
      if (head (temp).q2 == D.q1) then
        D := head (temp);
        pred := cons (D, pred)
      else
        temp := tail (temp)
      end if
    end loop
  end var
end function

------------------------------------------------------------------------

function delete_l_in_E (in var E: Edges, l: Street) : Edges is
  -- delete from E all edges with label l
  var new: Edges in
    new := {};
    loop
      case E
      var q1, q2: Corner, l1: Street in
        NIL ->
          return new
      | CONS (Edge (q1, l1, q2), E) ->
          if l1 != l then
            new := add_Edge (q1, q2, l1, new)
          end if
      end case
    end loop
  end var
end function

------------------------------------------------------------------------

function is_q1q2_in_E (in var E: Edges, A: Edge) : Bool is
  -- return true iff E has an edge with same source and target as A
  loop
    case E
    var q1, q2: Corner in
      NIL ->
        return false
    | CONS (Edge (q1, any Street, q2), E) ->
        if (q1 == A.q1) and (q2 == A.q2) then
          return true
        end if
    end case
  end loop
end function

------------------------------------------------------------------------

end module
\end{lstlisting}

\subsection{Definitions of AV specific datatypes}
\label{section:av-related}

Using the types and functions defined in module \lstinline+graph_car+ (see Section~\ref{section:graph_car}), the module \lstinline+types+ defines the AV specific datatypes together with their functions.
It also defines the channels, i.e., the types of the offers exchanged during rendezvous.

\begin{lstlisting}
module types (graph_car)
with ==, !=, get, set
is

------------------------------------------------------------------------
--  Types
------------------------------------------------------------------------

type Localization is !printedby "PRINT_LOCALIZATION"
  -- global localization (perception of the GPS), annotated with the
  -- current position of the car 
  Localization (c: Edge,  -- current street (i.e., edge) of the car
                G: Graph) -- current state of the global map)
end type

------------------------------------------------------------------------

type Radar_Grid is 
  -- local localization
  -- grid (obstacles or mvnt or nothing) (RADAR Grid)
  -- S0 initial states  
  Radar (E: Edges)
end type

------------------------------------------------------------------------

type Direction is 
  turn_n (N: Nat)  -- 5th if crossroads
end type

------------------------------------------------------------------------

type Itinerary is 
  -- instruction list  (-- rename  itinerary)
  list of Direction
with head, tail, reverse
end type

------------------------------------------------------------------------

type Control is 
  -- direction
  turned_n (N: Nat),
  brakes
end type

------------------------------------------------------------------------

type Obstacle is
  Obstacle (name: Street, position: Edge)
end type

------------------------------------------------------------------------

type Operation is
  -- obstacle operations
  random,
  turned_n (N: Nat),  -- 5th if crossroads,
  leave
end type

------------------------------------------------------------------------

type Obstacle_behaviour is
  list of Operation
with head, tail
end type

------------------------------------------------------------------------
--  Functions
------------------------------------------------------------------------

function succ_avoid_reach_D (E: Edges, P: Edge, D: Edge, avoid: Edges) 
: Edges is
  -- compute successors leading to D avoiding the edges in a given list
  var a: Edge, pred, succ, succ_t: Edges, dejavu: Vertices in
    if member (P, E) == false then
      return {}
    end if;
    a := P;
    pred := {a};
    succ := succ_l (E, a);
    succ_t := succ;
    dejavu := {a.q2};
    loop
      if member (D, succ_t) then
        return  pred_edge (pred, D) 
      elsif succ == {} then
        return {}
      else
        a := head (succ);
        succ := tail(succ);
        if (member (a.q2, dejavu) == false) and 
                    (member (a, avoid) == false) then
          succ_t := succ_l (E, a);
          pred := cons (a, pred);
          succ := union (succ_t, succ);
          dejavu := cons (a.q2, dejavu)
        end if
      end if
    end loop
  end var
end function

------------------------------------------------------------------------

function edges_to_itinerary (map: Localization, in var succ: Edges) 
: Itinerary is
  var eg: Edges, it: Itinerary, I: Nat in
    it := {};
    eg := map.G.E;
    loop
      if (succ == {}) or (eg == {}) then
        return reverse (it)
      end if;
      if head (eg) == head (succ) then
        I:= 0;
        succ := tail (succ);
        loop L in
          if (succ != {}) then
            eg := tail (eg);
            if eg == {} then
              eg := map.G.E
            end if;
            if head(eg).q1 == head (succ).q1 then
              if head(eg).q2 == head (succ).q2 then  
                it := cons (turn_n (I), it);
                break L
              end if;
              I := I + 1
            end if
          else
            return it
          end if
        end loop;
        eg := map.G.E
      else
        eg := tail (eg)
      end if
    end loop
  end var  
end function

------------------------------------------------------------------------

function compute_itinary (map: Graph, position, destination: Edge, 
                          avoid: Edges)
:  Itinerary is
  -- return if possible an itinerary leading from S to D in map
  var succ: Edges in
    succ := succ_avoid_reach_D (map.E, position, destination, avoid);
    return edges_to_itinerary (Localization (position, map), succ)
  end var
end function

------------------------------------------------------------------------

function car_controls (dir: Direction) : Control is
  case dir
  var N: Nat in
    turn_n (N) -> return turned_n (N)
  end case
end function

------------------------------------------------------------------------

function move_n (G: Graph, P: Edge, N: Nat) : Edge is
  -- return the car advancing in the n-th neighbourd
  var eg: Edges, i: Nat in
    eg := G.E;
    i := 0;
    loop L in
      if eg == {} then
        -- no successor found: do not move
        return P
      end if;
      if head (eg).q1 == P.q2 then
        if i == N then
          return head(eg)
        else
          i := i + 1
        end if
      end if;
      eg := tail (eg)
    end loop
  end var
end function

------------------------------------------------------------------------

function move_car (map: Localization, action: Control) : Localization is
  case action
  var N: Nat in
    turned_n (N) ->
      return Localization (move_n (map.G, map.c, N), map.G)
  | any ->
      return map
  end case
end function

------------------------------------------------------------------------

function move_obstacle (g: Graph, movement: Operation, o: Obstacle)
: Edge is
  case movement
  var N: Nat in
    turned_n (N) -> return move_n (g, o.position, N)
  | leave -> return Edge (0, HIDDEN,1)
  | random -> return Edge (0, HIDDEN,1)
  end case
end function

------------------------------------------------------------------------

function move_obstacle_grid (in var maj_radar: Radar_grid, 
                             obst: Obstacle) 
: Radar_Grid is 
  -- return the same obstacle position exists
  if is_q1q2_in_E (maj_radar.E, obst.position) then
    return maj_radar
  end if;
  -- delete the previous obstacle position
  maj_radar := Radar (delete_l_in_E (maj_radar.E, obst.name));
  -- add the new obstacle position
  return Radar (add_Edge (obst.position.q1, obst.position.q2, 
                          obst.name, maj_radar.E))
end function

------------------------------------------------------------------------

function succ_edge (in var R: Radar_Grid, D: Edge) 
: Radar_Grid is
  -- sucesseur of D
  var succ: Edges, temp: Edge in
    succ := {};
    loop L in
      if R.E == {} then
        return Radar (succ)
      end if;
      temp := head (R.E);
      if (temp.q1 == D.q2) then
        succ := cons (temp, succ)
      end if;
      R := R.{E -> tail (R.E)}
    end loop
  end var
end function

------------------------------------------------------------------------
--  Channels
------------------------------------------------------------------------
  
channel C is (c: Control) end channel
channel E is (e: Edges) end channel
channel G is (g: Radar_Grid) end channel
channel I is (i: Itinerary) end channel
channel O is (o: Obstacle) end channel
channel P is (e: Edge) end channel

------------------------------------------------------------------------

end module
\end{lstlisting}

\subsection{Definitions for the autonomous vehicle}
\label{section:car}

The four component of the autonomous car are defined as separate LNT processes \lstinline+PERCEPTION_RADAR+, \lstinline+PERCEPTION_GPS+, \lstinline+DECISION+, and \lstinline+ACTION+.
These four components are composed in parallel in the process \lstinline+CAR+ and synchronize on their shared actions using the parallel composition operator. 

\begin{lstlisting}[firstline=3,lastline=153]
------------------------------------------------------------------------

process PERCEPTION_RADAR [UPDATE_GRID, CURRENT_GRID: G] is
  -- local detection of obstacles
  var current, previous: Radar_Grid in
    UPDATE_GRID (?current);
    previous := current;
    CURRENT_GRID (current);
    loop
      -- receive the current grid
      UPDATE_GRID (?current);
      -- if the grid has changed, inform the driving controller
      if current != previous then
        CURRENT_GRID (current);
        previous := current
      end if
    end loop
  end var
end process

------------------------------------------------------------------------

process PERCEPTION_GPS [UPDATE_POSITION: P, 
                        REQUEST_POSITION: none, 
                        CURRENT_POSITION: P] is
  -- global localization
  var current_street: Edge, pending_request: Bool in
    -- initialize the car's position (i.e., the current street/edge)
    UPDATE_POSITION (?current_street);
    pending_request := false;
    loop
      select
        -- update the car's position (i.e., the current street/edge)
        UPDATE_POSITION (?current_street)
      []
        -- trajectory controller requests the position
        REQUEST_POSITION;
        pending_request := true
      []
        -- upon request, send the position to the trajectory controller 
        only if pending_request then
          CURRENT_POSITION (current_street);
          pending_request := false
        end if
      end select
    end loop
  end var
end process

------------------------------------------------------------------------

process DECISION [REQUEST_PATH: E,
                  CURRENT_PATH: I,
                  REQUEST_POSITION: none,
                  CURRENT_POSITION: P,
                  ARRIVAL: none]
                 (map: Graph, destination: Edge) is
  var l_avoid: Edges, path: Itinerary, current_street: Edge in
    loop
      -- wait for a request from the driving controller
      REQUEST_PATH (?l_avoid);
      -- request the current position
      REQUEST_POSITION;
      CURRENT_POSITION (?current_street);
      if current_street == destination then
        -- the car arrived to the final destination
        ARRIVAL;
        -- stop
        loop i end loop
      end if;
      -- compute an itineray and send it to the driving controller
      path := compute_itinary (map, current_street, destination,
                               l_avoid);
      CURRENT_PATH (path)
    end loop
  end var
end process

------------------------------------------------------------------------

process ACTION [REQUEST_PATH: E,
                CURRENT_PATH: I,
                CURRENT_GRID: G,
                CAR_MOVE: C,
                COLLISION: P] is
  var grid: Radar_Grid, path: Itinerary in
    grid := Radar ({});
    loop
      -- check feasibility of the itinerary
      REQUEST_PATH (grid.E);
      CURRENT_PATH (?path);
      if path == {} then
        -- wait for the obstacles to move ...
        CURRENT_GRID (?grid)
      else
        loop L in
          select
            if path == {} then -- last direction sent
              break L
            else
              CAR_MOVE (car_controls (head (path)));
              path := tail (path)
            end if
          []
            -- receive the current local grid from the radar
            -- this branch should be prior than the previous one to
            -- never block the reception of a new grid from the radar
            CURRENT_GRID (?grid);
            break L
          []
            COLLISION (?any Edge);
            loop i end loop
          end select
        end loop
      end if
    end loop
  end var
end process

------------------------------------------------------------------------

process CAR [UPDATE_GRID: G,
             UPDATE_POSITION: P,
             CURRENT_GRID: G,
             REQUEST_POSITION: none,
             CURRENT_POSITION: P,
             REQUEST_PATH: E,
             CURRENT_PATH: I,
             CAR_MOVE: C,
             ARRIVAL: none,
             COLLISION: P]
            (map: Graph, destination: Edge) is
  par
    CURRENT_GRID ->
      PERCEPTION_RADAR [UPDATE_GRID, CURRENT_GRID]
  ||
    REQUEST_POSITION, CURRENT_POSITION ->
      PERCEPTION_GPS [UPDATE_POSITION, REQUEST_POSITION,
                      CURRENT_POSITION]
  ||
    REQUEST_PATH, CURRENT_PATH, REQUEST_POSITION, CURRENT_POSITION ->
      DECISION [REQUEST_PATH, CURRENT_PATH, REQUEST_POSITION,
                CURRENT_POSITION, ARRIVAL] (map, destination)
  ||
    CURRENT_PATH, CURRENT_GRID, REQUEST_PATH ->
      ACTION [REQUEST_PATH, CURRENT_PATH, CURRENT_GRID, CAR_MOVE,
              COLLISION]
  end par
end process

------------------------------------------------------------------------
\end{lstlisting}

\subsection{Definitions for the autonomous vehicles environment}
\label{section:environment}

The environment of the car is modeled by two LNT processes.
First, the process \lstinline+OBSTACLE+ specifies the behaviour of an obstacle.
Second, the process \lstinline+MAP_MANAGEMENT+ manages the ground truth, i.e., the positions of the car and obstacles on the map. 
Because many obstacles can be interacting with the AV, we added a third process \lstinline+OBSTACLES_MANAGER+ instantiating the number of obstacles, this process is presented in the next section. 
The \lstinline+MAP_MANAGEMENT+ and \lstinline+OBSTACLES_MANAGER+ processes are composed in the process \lstinline+Environment+.

\begin{lstlisting}[firstline=152,lastline=295]

------------------------------------------------------------------------

process OBSTACLE [CAR_POSITION: P, OBSTACLES_GRID: G, OBSTACLE_MOVE: O]
                 (map: Localization,
                  o: Obstacle,
                  in var operations: Obstacle_behaviour) is
  -- execute the sequence of operations and then stop
  var present: Bool, grid: Radar_grid, car_street: Edge in
    present := false;
    grid := Radar ({});
    car_street := map.c;
    loop
      select
        -- apparition of the obstacle at its initial position
        only if not (present or
                     is_q1q2_in_E (cons (car_street, grid.E), 
                                   o.position))
        then
          if member (o.position, map.G.E) then
            OBSTACLE_MOVE (o);
            present := true
          end if 
        end if
      []
        -- movement of the obstacle
        only if present and (operations != {}) then
          var pos: Edge, opi: Operation in
            opi := head (operations);
            if opi == random then
              select
                opi := leave
              []
                var n0, n1: Nat in
                  n1 := length (succ_l (map.G.E, o.position));
                  n0 := any Nat where n0 <= n1;
                  opi := turned_n (n0)
                end var
              end select
            end if;
            pos := move_obstacle (map.G, opi, o);
            if is_q1q2_in_E (cons (car_street, grid.E), 
                             pos) == false
            then
              OBSTACLE_MOVE (Obstacle (o.name, pos));
              operations := tail (operations)
            else
              i -- to avoid problems with the translation to LOTOS
            end if
          end var
        end if
      []
        -- update the map with the current street of the car
        CAR_POSITION (?car_street)
      [] 
        -- update the grid with the current position of the obstacles
        OBSTACLES_GRID (?grid)
      end select
    end loop
  end var
end process

------------------------------------------------------------------------

process MAP_MANAGEMENT [UPDATE_POSITION: P,
                        UPDATE_GRID: G,
                        CAR_POSITION: P,
                        OBSTACLES_GRID: G,
                        CAR_MOVE: C,
                        OBSTACLE_MOVE: O,
                        COLLISION: P] 
                       (in global_loc: Localization) is

  var map: Localization, grid: Radar_Grid, car_control: Control in
    map := global_loc;
    grid := Radar ({});
    -- initialize the position of the car: the map in the GPS does not
    -- change, thus send only the current street
    UPDATE_POSITION (map.c);
    loop
      select
        -- handle car movement
        CAR_MOVE (?car_control);
        -- compute car position and new grid for the radar
        map := move_car (map, car_control);
        -- check if the car is in the same edge as an obstacle
        if is_q1q2_in_E (grid.E, map.c) then
          COLLISION (map.C);
          -- stop
          loop i end loop
        end if;
        -- inform the GPS, the radar, and the obstacles
        -- as before, no need for a parallel composition
        UPDATE_POSITION (map.c);
        CAR_POSITION (map.c);
        UPDATE_GRID (succ_edge(grid, map.c))
      []
        var obst: Obstacle in 
          -- handle obstacle movement
          OBSTACLE_MOVE (?obst);
          -- compute new grid for the radar
          grid := move_obstacle_grid (grid, obst);
          -- inform the radar and the obstacles about the change
          -- no need for a parallel composition, as long as the
          -- environment itself is sequential
          UPDATE_GRID (succ_edge(grid, map.c));
          OBSTACLES_GRID (grid)
        end var
      end select
    end loop
  end var
end process

------------------------------------------------------------------------

process ENVIRONMENT [UPDATE_POSITION: P,
                     UPDATE_GRID: G,
                     CAR_MOVE: C,
                     OBSTACLE_MOVE: O, 
                     ARRIVAL: none,
                     COLLISION: P] 
                    (in global_loc: Localization) is
  disrupt
    hide CAR_POSITION: P, OBSTACLES_GRID: G in
      par OBSTACLE_MOVE, CAR_POSITION, OBSTACLES_GRID in
        OBSTACLES_MANAGER [CAR_POSITION, OBSTACLES_GRID, OBSTACLE_MOVE]
                             (global_loc)
      ||
        MAP_MANAGEMENT [UPDATE_POSITION, UPDATE_GRID, 
                        CAR_POSITION, OBSTACLES_GRID,
                        CAR_MOVE,
                        OBSTACLE_MOVE,
                        COLLISION] 
                       (global_loc)
      end par
    end hide               
  by
    -- stop the environment, when the car arrives at the destination
    ARRIVAL;
    loop i end loop -- livelock to avoid warnings from the LNT compiler
  end disrupt
end process

------------------------------------------------------------------------
\end{lstlisting}

The processes described in Sections~\ref{section:car} and \ref{section:environment} are grouped into a module \lstinline+car+, collecting the behavioral specification.

\subsection{Example of a simulation scenario (scene and actor behaviour)}

A scenario, or specific instance of the model, is characterized by the map, together with the initial position and destination of the car, and the set of obstacles with their initial position and behavior.
This information is grouped in the \lstinline+main+ module, providing the (constant) functions for the \lstinline+initial_map+, the process handling the parallel composition of the various obstacles, and the principal process \lstinline+MAIN+ instantiating everything. 

\begin{lstlisting}
module main (car) is

------------------------------------------------------------------------

function initial_map : Graph is
  var g: Graph, e: Edges, v: Vertices in
    v := {0, 1, 2, 3, 4, 5, 6, 7, 8};
    -- ATTENTION: the list e must be ordered by increasing source states
    e := {Edge (0, Coronation_Street, 1), 
          Edge (0, Corporation_Street, 3), 
          Edge (1, Coronation_Street_bis, 0),
          Edge (1, two_Coronation_Street, 2),
          Edge (1, Sackville, 4),
          Edge (2, two_Coronation_Street_bis, 1), 
          Edge (2, Spring_Gardens, 5),
          Edge (3, Corporation_Street_bis, 0),
          Edge (3, Princess_Street, 4),
          Edge (3, two_Corporation_Street, 6),
          Edge (4, two_Princess_Street, 5),
          Edge (4, two_Sackville, 7),
          Edge (5, Spring_Gardens_bis, 2),
          Edge (5, two_Princess_Street_bis, 4),
          Edge (5, two_Spring_Gardens, 8),
          Edge (6, two_Corporation_Street_bis, 3),
          Edge (6, New_Cathedral_Street, 7),
          Edge (7, two_Sackville_bis, 4),
          Edge (7, New_Cathedral_Street_bis, 6),
          Edge (7, two_New_Cathedral_Street, 8),
          Edge (8, two_Spring_Gardens_bis, 5),
          Edge (8, two_New_Cathedral_Street_bis, 7)
          };
    g := Graph (v, e);
    return g
  end var
end function 

------------------------------------------------------------------------

process OBSTACLES_MANAGER [CAR_POSITION: P,
                           OBSTACLES_GRID: G,
                           OBSTACLE_MOVE: O]
                             (map: Localization) is
  var baby, cyclist: Obstacle, b1, b2: Obstacle_behaviour in
    -- initialisation of obstacle position
    baby := Obstacle (Lily, Edge (1, Sackville, 4));
    cyclist := Obstacle (Theo, Edge (5, two_Princess_Street_bis, 4));

    -- initialisation of obstacle behaviours
    b1 := {random};
    b2 := {random};
    
    par CAR_POSITION, OBSTACLES_GRID in
      Obstacle [CAR_POSITION, OBSTACLES_GRID, OBSTACLE_MOVE] 
              (map, baby, b1)
    ||
      Obstacle [CAR_POSITION, OBSTACLES_GRID, OBSTACLE_MOVE] 
              (map, cyclist, b2)
    end par
  end var
end process

------------------------------------------------------------------------

process MAIN [UPDATE_GRID: G,
              UPDATE_POSITION: P,
              CURRENT_GRID: G,
              REQUEST_POSITION: none,
              CURRENT_POSITION: P,
              REQUEST_PATH: E,
              CURRENT_PATH: I,
              CAR_MOVE: C,
              OBSTACLE_MOVE: O,
              ARRIVAL: none,
              COLLISION: P,
              raise BAD_PARAMETERS: none]
             (source, destination: Street) is
  if not (is_l_in_E (initial_map.E, source) and 
     is_l_in_E (initial_map.E, destination)) then
       raise BAD_PARAMETERS
  end if;
  par UPDATE_GRID, UPDATE_POSITION, CAR_MOVE, ARRIVAL, COLLISION in
    CAR [UPDATE_GRID, UPDATE_POSITION,
         CURRENT_GRID, REQUEST_POSITION, CURRENT_POSITION, 
         REQUEST_PATH, CURRENT_PATH, CAR_MOVE,
         ARRIVAL, COLLISION]
        (initial_map, edge_l (initial_map.E, destination))
  ||
    ENVIRONMENT [UPDATE_POSITION, UPDATE_GRID,
                 CAR_MOVE, OBSTACLE_MOVE, ARRIVAL, COLLISION] 
                (Localization (edge_l (initial_map.E, source),
                               initial_map))
  end par
end process

------------------------------------------------------------------------

end module
\end{lstlisting}

\section{Full model focused on perception}
\label{ap:grid-model}

\tikzset{
    artifact/.style={draw,trapezium,trapezium left angle=82,trapezium right angle=98,text width=1.5cm,align=center,font=\small\sffamily}
}

\begin{figure}
    \centering
\scalebox{.80}{
    \begin{tikzpicture}[x=2.25cm,y=0.9cm]
    %---
    % Borders
    %\draw (-0.87,4) rectangle (9,0);
    % Boxes: 
    \node[artifact] (types) at (-0.16,2.8) {Types};
    \node[artifact, above right = 12pt of types] (map) {Map};
    \node[artifact, below right = 12pt of types] (lidar) {LIDAR};
    \node[artifact, right = 12pt of map] (obst) {Obstacles};
    \node[artifact, right = 12 pt of lidar] (car) {Car};
    \node[artifact, below right = 12pt of obst] (scenario) {Scenario};
    \node[artifact, right = 12pt of scenario] (mapm) {Map \\ Manager};
    \node[artifact, right = 12pt of mapm] (main) {Main};
    
    %---
    \begin{scope}[every path/.style={-latex}]
    \draw [line width=1pt] (types) edge (map)
                           (types) edge (lidar)
                           (map) edge (obst)
                           (lidar) edge (car)
                           (obst) edge (scenario)
                           (car) edge (scenario)
                           (scenario) edge (mapm)
                           (mapm) edge (main);
    \end{scope}
    \end{tikzpicture}
}
\vspace{-1ex}

    \caption{Hierarchy of the LNT modules}
    \label{fig:model_architecture}
\end{figure}
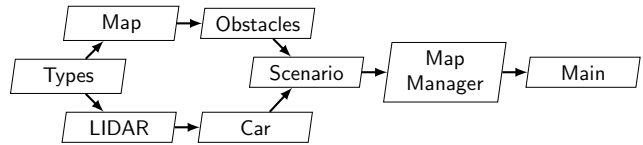

This appendix contains the full LNT code of the model with the refined array-based representation of the map required by the focus on the perception.
There is a separate subsection for each module.
The order of subsections follows the module inclusion as shown in architecture depicted in Figure~\ref{fig:model_architecture} (from the deepest inclusion to the top level module). 

\subsection{Definitions of datatypes}

Module \lstinline+modele_types+ defines the constants (including the size of the arrays representing the map and the perception grid), the data types encoding the different actors and their attributes, and the channels used for communication between processes, together with two extra functions used to factor the update of the position for the obstacle data type.

\begin{lstlisting}
module modele_types
with ==, !=, get, set
is

------------------------------------------------------------------------
-- Constants
------------------------------------------------------------------------

-- constant defining the height of the array representing the map
function max_map_height : Nat is
  return 9
end function

-- same constant as an Int (rather than a Nat)
function max_map_height_int : Int is
  return NatToInt (max_map_height)
end function

------------------------------------------------------------------------

-- constant defining the width of the array representing the map
function max_map_width : Nat is
  return 9
end function

-- same constant as an Int (rather than a Nat)
function max_map_width_int : Int is
  return NatToInt (max_map_width)
end function

------------------------------------------------------------------------

-- constant defining the size of the square array representing the LiDAR
function max_lidar_size : Nat is
  return 4
end function

------------------------------------------------------------------------

-- constant given as coordinate to an obstacle outside of the map array
-- note: must be greater than both max_map_height and max_map_width
function out_of_array : Nat is
  return 50
end function

-- same constant as an Int (rather than a Nat)
function out_of_array_int : Int is
  return NatToInt (out_of_array)
end function

------------------------------------------------------------------------
-- Types
------------------------------------------------------------------------

-- name of the obstacles
type Obstacle_name is
  Pedestrian,
  PedestrianB,
  Other_car,
  Other_carB,
  Cyclist,
  Scenery,
  Misc
end type

------------------------------------------------------------------------

-- possible value of the map's cells
type Map_Info is 
  occupied (O:Obstacle),
  car_pos,
  free
end type

------------------------------------------------------------------------

-- one line of the map
type Map_line is
  array [0..9 (* max_map_width *)] of Map_Info
end type

------------------------------------------------------------------------

-- the map (ground truth)
type Map is
  array [0..9 (* max_map_height *)] of Map_line
end type

------------------------------------------------------------------------

-- Possible value of the perception's cells
type Perception_Info is 
  C, -- car position
  F, -- free 
  O, -- occupied by an obstacle
  M, -- occupied by a moving obstacle
  T, -- occupied by a transparent obstacle
  N, -- occupied by a moving transparent obstacle
  U  -- unknown
end type

------------------------------------------------------------------------

-- one line of the perception grid
type L (* Perception_Line *) is
  array [0..4 (* max_lidar_size *)] of Perception_Info
end type

------------------------------------------------------------------------

-- the perception grid
type P (* Perception *) is
  array [0..4 (* max_lidar_size *)] of L
end type

------------------------------------------------------------------------

-- position of an element on the map; x column, y line
-- note: Int rather than Nat to allow actors to leave the map 
type Position is
  Position (x: Int, y: Int)
end type

------------------------------------------------------------------------

-- position of a rectangular object (upper left and bottom right corner)
type RectPosition is
  RectPosition (pos1: Position, pos2: Position)
end type

------------------------------------------------------------------------

-- number of cells an obstacle can travel in a single move
type Obstacle_Speed is 
  -- static obstacles (speed = 0)
  range 0..3 of NAT
with <, <=, >, >=, first, last
end type

------------------------------------------------------------------------

-- directions in which an actor can move, one random direction
type Direction is
  -- moves to adjacent cells
  up,
  down,
  right,
  left,
  -- diagonal moves
  upright,
  upleft,
  downright,
  downleft,
  -- particular moves
  none,
  random
  with ==
end type

------------------------------------------------------------------------

-- list of directions defining an actor behavior
type Directions is
!card 35
  list of Direction
with head, tail, reverse, length 
end type

------------------------------------------------------------------------

-- obstacle, charaterized by a name, a rectangular position, a speed
-- an initial direction, and whether it is transparent or not
type Obstacle is
  Obstacle (name: Obstacle_name,
            position: RectPosition, 
            speed: Obstacle_Speed,
            dir: Direction,
            transparent: Bool)
end type

------------------------------------------------------------------------
-- Channels
------------------------------------------------------------------------

channel C is
  (current: Position, previous: Position)
end channel

channel G is
  (ground_truth: Map)
end channel

channel O is
  (current: Obstacle, previous: Obstacle, move: Direction)
end channel

channel L is
  (perception: P, ground_truth: Map)
end channel

channel CG is
  (current: Position, ground_truth: Map)
end channel

channel COLLISION_T is
  (collision_at: Position)
end channel

channel COLLISION_O is
  (collision_with: Obstacle_name)
end channel

------------------------------------------------------------------------
--  Functions
------------------------------------------------------------------------

-- wrapper of + for type Obstacle_Speed
function + (s1, s2: Obstacle_Speed) : Obstacle_Speed is
  return Obstacle_Speed (Nat (s1) + Nat (s2))
end function

------------------------------------------------------------------------

-- return the value of the cell of the array m in the coordinates (x, y)
function cell_value (m: Map, x, y: Int) : Map_Info is
  return m[IntToNat (y)][IntToNat (x)]
end function

------------------------------------------------------------------------

-- function transformg a Position into a single-cell RectPosition
function CellRectPosition (pos: Position) : RectPosition is
  return RectPosition (pos, pos)
end function

------------------------------------------------------------------------

-- return the distance (computed as the sum of the squared differences
-- between coordinates) between two positions
function dist (pos1, pos2: position) : Int is
  var x, y: Int in
    x := pos2.x - pos1.x;
    y := pos2.y - pos1.y;
    return (x * x) + (y * y)
  end var
end function

------------------------------------------------------------------------

-- return true iff position (x, y) is free in m
function free (m: Map, x, y: Int) : Bool is
  return cell_value (m, x, y) == free
end function

------------------------------------------------------------------------

-- update map m by setting contents of cell (x, y) to v
function update_cell (in out m: Map, x, y: Int, v: Map_Info) is
  var l: Map_line in
    l := m[IntToNat (y)];
    l[IntToNat (x)] := v;
    m[IntToNat (y)] := l
  end var
end function

------------------------------------------------------------------------

-- update the parameter x for a Position
function update_position_x (pos: Position, N: Int) : Position is
  return pos.{x -> N}
end function

------------------------------------------------------------------------

-- update the parameter y for a Position
function update_position_y (pos: Position, N: Int) : Position is
  return pos.{y -> N}
end function

------------------------------------------------------------------------

-- update the parameter x for a RectPosition
-- N1 is for the top left corner, N2 for the bottom right one
function update_rectpos_x (in var pos: RectPosition, N1, N2: Int)
: RectPosition is
  pos := pos.{pos1 -> pos.pos1.{x -> N1}};
  return pos.{pos2 -> pos.pos2.{x -> N2}}
end function

------------------------------------------------------------------------

-- update the parameter y for a RectPosition
-- N1 is for the top left corner, N2 for the bottom right one
function update_rectpos_y (in var pos: RectPosition, N1, N2: Int)
: RectPosition is
  pos := pos.{pos1 -> pos.pos1.{y -> N1}};
  return pos.{pos2 -> pos.pos2.{y -> N2}}
end function

------------------------------------------------------------------------

end module
\end{lstlisting}

\subsection{Definition of the LiDAR}

Module \lstinline+modele_lidar+ contains the functions to create and modify the perception grid, together with the process managing the grid.

\begin{lstlisting}
module modele_lidar (modele_types) is

------------------------------------------------------------------------
--  Functions
------------------------------------------------------------------------

-- convert the contents of a cell of the ground truth into the contents
-- of the corresponding cell of the perception grid
function array_to_perception (cell: Map_Info) : Perception_Info is
  case cell in
    occupied (any Obstacle) ->
      if cell.O.transparent then
        return T
      else
        return O
      end if
  | car_pos ->
      return C
  | free ->
      return F
  end case
end function

------------------------------------------------------------------------

-- shorthand to access cell map[y+j][x+i]
function map_value (map: Map, x, y: Int, i, j: Nat) : Map_Info is
  return map [IntToNat (y + NatToInt (j))][IntToNat (x + NatToInt (i))]
end function

------------------------------------------------------------------------

-- change the LiDAR cell in the position (x, y) to the perception info v
function update_lidar (in out lid: P, x, y: Nat, v: Perception_Info) is
  var b: L in
    b := lid[y];
    b[x] := v;
    lid[y] := b
  end var
end function

------------------------------------------------------------------------

-- change the LiDAR grid by setting to U all cells hidden by a non
-- transparent obstacle at position (x, y) ; the positions of the
-- cells to hide are (a, b), (c, d), and (e, f)
function hide_cells (in out lid: P, x, y, a, b, c, d, e, f: Nat) is
  if lid[y][x] == O then 
    update_lidar (!?lid, a, b, U);
    update_lidar (!?lid, c, d, U);
    update_lidar (!?lid, e, f, U)
  end if
end function

------------------------------------------------------------------------

-- compare two LiDAR grids lid and prev_lid, and change any O to M and
-- T to N in lid, if the cells are free in prev_lid
function report_moving_cells (in out lid: P, prev_lid: P) is
  var I, J: Nat in
    I := 0;
    while I <= max_lidar_size loop
      J := 0;
      while J <= max_lidar_size loop
        if prev_lid[J][I] == F then
          if lid[J][I] == O then
            update_lidar (!?lid, I, J, M)
          else 
            if lid[J][I] == T then
              update_lidar (!?lid, I, J, N)
            end if
          end if
        end if;
        J := J + 1
      end loop;
      I := I + 1
    end loop
  end var
end function

------------------------------------------------------------------------

-- create a perception grid depending on the position of the car and the
-- elements around it, and hide the cells behind any non transparent
-- obstacle
function fill_lidar (car: Position, map: Map, prev:P) : P is
  var newLid: P, X, Y: Int, I, J: Nat in
    X := car.x - 2;
    Y := car.y - 2;
    newLid := P (L (F)); -- initially, all cells are considered free
    I := 0;
    while I <= max_lidar_size loop
      J := 0;
      while J <= max_lidar_size loop
        if (Y + NatToInt (J) > max_map_height_int) or
           (X + NatToInt (I) > max_map_width_int) or
           (Y + NatToInt (J) < 0) or
           (X + NatToInt (I) < 0) then
          -- car cannot perceive beyond the edge of the map
          update_lidar (!?newLid, I, J, U)
        else
          update_lidar (!?newLid, I, J,
                     array_to_perception (map_value (map, X, Y, I, J)))
        end if;
        J := J + 1
      end loop;
      I := I + 1
    end loop;
    -- check the cells around the car and compute appropriate hiding
    hide_cells (!?newLid, 1, 1, 0, 0, 0, 1, 1, 0);
    hide_cells (!?newLid, 1, 2, 0, 1, 0, 2, 0, 3);
    hide_cells (!?newLid, 1, 3, 0, 3, 0, 4, 1, 4);
    hide_cells (!?newLid, 2, 1, 1, 0, 2, 0, 3, 0);
    hide_cells (!?newLid, 2, 3, 1, 4, 2, 4, 3, 4);
    hide_cells (!?newLid, 3, 1, 3, 0, 4, 0, 4, 1);
    hide_cells (!?newLid, 3, 2, 4, 1, 4, 2, 4, 3);
    hide_cells (!?newLid, 3, 3, 3, 4, 4, 3, 4, 4);
    report_moving_cells (!?newLid, prev);
    return newLid
  end var
end function

------------------------------------------------------------------------
--  Processes
------------------------------------------------------------------------

-- process updating the LiDAR depending one the position of the car 
-- and the state of the map
process LIDAR_MANAGER [GRID_CAR:CG, LIDAR_MAP: L] (updateLidar: Bool) is
  var lid, prev_lid: P, grid : Map, car: Position in
    lid := P (L (F));
    loop
      prev_lid := lid;
      GRID_CAR (?car, ?grid);
      if updateLidar then
        lid := fill_lidar (car, grid, prev_lid)
      end if;
      LIDAR_MAP (lid, grid)
    end loop
  end var
end process

------------------------------------------------------------------------

end module
\end{lstlisting}

\subsection{Definitions of the car}

Module \lstinline+modele_car+ contains the functions to compute the moves of the car and the process managing the car's behaviour, according to the current configuration of the ground truth and the arguments given in the scenario module to define the car's behaviour.

\begin{lstlisting}
module modele_car (modele_lidar) is

------------------------------------------------------------------------
-- Functions
------------------------------------------------------------------------

-- function to move the car c in the fixed direction dir for a number of
-- steps corresponding to speed ; the move stops earlier if it traverses
-- a non free cell ; diagonal moves move one cell at a time to allow
-- different turn ranges independent from speed
function car_move (in out c: Position, dir: Direction, grid: Map,
                   speed: Nat) is
  var N, N2: Int, S: Nat in
    S := 0;
    case dir in
      up -> -- y-1
        loop L in
          if ((c.y == 0) or (S == speed)) then 
            break L
          end if;
          N := c.y - 1;
          c := update_position_y (c, N);
          if not (free (grid, c.x, c.y)) then
            break L
          end if;
          S := S + 1
        end loop
    | down -> -- y+1
        loop L in
          if (c.y == max_map_height_int) or (S == speed) then 
            break L
          end if;
          N := c.y + 1;
          c := update_position_y (c, N);
          if not (free (grid, c.x, c.y)) then
            break L
          end if;
          S := S + 1
        end loop
    | right -> -- x+1
        loop L in
          if (c.x == max_map_width_int) or (S == speed) then 
            break L
          end if;
          N := c.x + 1;
          c := update_position_x (c, N);
          if not (free (grid, c.x, c.y)) then
            break L
          end if;
          S := S + 1
        end loop
    | left -> -- x-1
        loop L in
          if (c.x == 0) or (S == speed) then 
            break L
          end if;
          N := c.x - 1;
          c := update_position_x (c, N);
          if not (free (grid, c.x, c.y)) then
            break L
          end if;
          S := S + 1
        end loop
    | upright -> -- x+1 y-1
        if not ((c.x == max_map_width_int) or (c.y == 0)) then 
          N := c.x + 1;
          N2 := c.y - 1;
          c := update_position_x (c, N);
          c := update_position_y (c, N2)
        end if
    | upleft -> -- x-1 y-1
        if not ((c.x == 0) or (c.y == 0)) then 
          N := c.x - 1;
          N2 := c.y - 1;
          c := update_position_x (c, N);
          c := update_position_y (c, N2)
        end if
    | downright -> -- x+1 y+1
        if not ((c.x == max_map_width_int) or 
                (c.y == max_map_height_int)) then 
          N := c.x + 1;
          N2 := c.y + 1;
          c := update_position_x (c, N);
          c := update_position_y (c, N2)
        end if
    | downleft -> -- x-1 y+1
        if not ((c.x == 0) or (c.y == max_map_height_int)) then 
          N := c.x - 1;
          N2 := c.y + 1;
          c := update_position_x (c, N);
          c := update_position_y (c, N2)
        end if
    | any -> null -- never reached
    end case
  end var
end function

------------------------------------------------------------------------
-- Processes
------------------------------------------------------------------------

-- animate the car according to behaviour
process MOVE_CAR [LIDAR_MAP:L, CAR_POSITION:C, ARRIVAL: none]
                 (in var pos: Position,
		  in var behaviour: Directions,
		  cycle: Bool,
		  speed: Nat)
is
  var initial_behaviour: Directions, prev_pos: Position, grid: Map in
    initial_behaviour := behaviour;
    prev_pos := pos;
    select
      null
    []
      LIDAR_MAP (?any P, ?any Map)
    end select;
    CAR_POSITION (pos, prev_pos);
    loop
      LIDAR_MAP (?any P, ?grid);
      select
        null
      []
	if behaviour == {} then
	  if cycle then
	    behaviour := initial_behaviour
	  else
            ARRIVAL;
            loop i end loop
	  end if
        end if;
        prev_pos := pos;
        eval car_move (!?pos, head (behaviour), grid, speed);
        CAR_POSITION (pos, prev_pos);
        behaviour := tail (behaviour)
      end select
    end loop
  end var
end process

------------------------------------------------------------------------

end module
\end{lstlisting}

\subsection{Definitions of the geographical map}

Module \lstinline+modele_map+ contains all necessary functions to manage the array representing the ground truth.
These functions are used to update the array and check the value of cells when moving an actor.
The module also contains functions factoring some features, such as the update of an attribute or the access to an array cell.

\begin{lstlisting}
module modele_map (modele_types) is

------------------------------------------------------------------------
--  Functions
------------------------------------------------------------------------

-- update the position of an obstacle for the x coordinates
function obst_update_position_x (in out obst: Obstacle, N1, N2: Int) is
  obst := obst.{position -> update_rectpos_x (obst.position, N1, N2)}
end function

------------------------------------------------------------------------

-- update the position of an obstacle for the y coordinates
function obst_update_position_y (in out obst: Obstacle, N1, N2: Int) is
  obst := obst.{position -> update_rectpos_y (obst.position, N1, N2)}
end function

------------------------------------------------------------------------

-- update all coordinates of an obstacle out of the map to out_of_array
function obst_update_position (in out obst: Obstacle) is
  if obst.position.pos1.x == out_of_array_int then
    -- by construction, all other coordinates of O are also out_of_array
    null -- nothing to do
  else
    obst_update_position_x (!?obst, out_of_array_int, out_of_array_int);
    obst_update_position_y (!?obst, out_of_array_int, out_of_array_int)
  end if
end function

------------------------------------------------------------------------

-- return true iff the coordinates (x, y) are outside the ground truth
function is_outside (x, y: Int) : Bool is
  return (x > max_map_width_int)  or (x < 0) or
         (y > max_map_height_int) or (y < 0)
end function

------------------------------------------------------------------------

-- return true iff obstacle obst is completely out of the ground truth
function is_obstacle_out (obst: Obstacle) : Bool is
  return is_outside (obst.position.pos1.x, obst.position.pos1.y) and
         is_outside (obst.position.pos2.x, obst.position.pos2.y)
end function

------------------------------------------------------------------------

-- add obstacle obst in the right position on the ground truth array and
-- return the updated grid, checking whether the obstacle is at least
-- partially in the array
function add_obstacle (in out m: Map, obst: Obstacle) is
  if obst.position.pos1 == obst.position.pos2 then
    -- special case of a single-cell obstacle
    update_cell (!?m, obst.position.pos1.x, obst.position.pos1.y,
                 Occupied (obst))
  else
    var x, y: Int in
      x := obst.position.pos1.x;
      while x <= obst.position.pos2.x loop
        y := obst.position.pos1.y;
        while y <= obst.position.pos2.y loop
          if not (is_outside (x, y)) then
            update_cell (!?m, x, y, Occupied(obst))
          end if;
          y := y + 1
        end loop;
        x := x + 1
      end loop
    end var
  end if
end function

------------------------------------------------------------------------

-- remove obstacle obst from the ground truth array m
function remove_obstacle (in out m: Map, obst: Obstacle) is
  if obst.position.pos1 == obst.position.pos2 then
    -- special case of a single-cell obstacle
    update_cell (!?m, obst.position.pos1.x, obst.position.pos1.y, free)
  else
    var x, y: Int in
      x := obst.position.pos1.x;
      while x <= obst.position.pos2.x loop
        y := obst.position.pos1.y;
        while y <= obst.position.pos2.y loop
          if not (is_outside (x, y)) then
            update_cell (!?m, x, y, free)
          end if;
          y := y + 1
        end loop;
        x := x + 1
      end loop
    end var
  end if
end function

------------------------------------------------------------------------

-- for an obstacle O at RectPosition p moving in direction dir, return
-- true iff at least one cell of the next RectPosition occupied by O
-- after the move is not free
-- note: currently, diagonal moves are only supported for single-cell
-- obstacles
function is_next_pos_occupied (p: RectPosition, dir: Direction, m: Map)
: Bool is
  var x, y: Int in
    case dir in
      up ->
        y := p.pos1.y - 1;
        if y >= 0 then
          for x := p.pos1.x while x <= p.pos2.x by x := x + 1 loop
            if not (free (m, x, y)) then
              return true
            end if
          end loop
        end if;
        return false
    | down ->
        y := p.pos1.y - 1;
        if y <= max_map_height_int then
          for x := p.pos1.x while x <= p.pos2.x by x := x + 1 loop
            if not (free (m, x, y)) then
              return true
            end if
          end loop
        end if;
        return false
    | right ->
        x := p.pos1.x + 1;
        if x <= max_map_width_int then
          for y := p.pos1.y while y <= p.pos2.y by y := y + 1 loop
            if not (free (m, x, y)) then
              return true
            end if
          end loop
        end if;
        return false
    | left ->
        x := p.pos1.x - 1;
        if x >= 0 then
          for y := p.pos1.y while y <= p.pos2.y by y := y + 1 loop
            if not (free (m, x, y)) then
              return true
            end if
          end loop
        end if;
        return false
    | upright ->
        return (p.pos1.x + 1 <= max_map_width_int) and 
               (p.pos1.y - 1 >= 0) and
               not (free (m, p.pos1.x + 1, p.pos1.y - 1))
    | upleft ->
        return (p.pos1.x - 1 >= 0) and
               (p.pos1.y - 1 >= 0) and
               not (free (m, p.pos1.x - 1, p.pos1.y - 1))
    | downleft ->
        return (p.pos2.x - 1 >= 0) and
               (p.pos2.y + 1 <= max_map_height_int) and
               not (free (m, p.pos2.x - 1, p.pos2.y + 1))
    | downright ->
        return (p.pos2.x + 1 <= max_map_width_int) and
               (p.pos2.y + 1 <= max_map_height_int) and
               not (free (m, p.pos2.x + 1, p.pos2.y + 1))
    | none ->
        -- the obstacle already occupies the cells
        return false
    | any ->
        return false -- never reached
    end case
  end var
end function

------------------------------------------------------------------------

end module
\end{lstlisting}

\subsection{Definitions of the obstacles}

Module \lstinline+modele_obstacles+ contains the function to correctly compute obstacle moves and the process managing the behaviour of the obstacles, depending on the arguments given in the scenario module (see Section~\ref{section:modele_scenario} and the current version of the grid received via the gate with channel \lstinline+G+. It also contains the functions used to restrain the random moves of an obstacle.

\lstinputlisting{modele_obstacles.lnt}

\subsection{Example of a simulation scenario (scene and actor behaviour)}
\label{section:modele_scenario}

Module \lstinline+modele_scenario+ is an example of a scenario definition.
It illustrates how the model is configured for a particular scenario (scenery and actors with their behaviour).
Concretely, module \lstinline+modele_scenario+ defines a ten$\times$ten ground truth array scenery with four buildings (i.e., a X-shaped crossroad as shown in Figure~\ref{fig:map_grid} (x-coordinates increase from zero on the left to ten on the right, y-coordinates increase from zero on top to ten on the bottom).
Moving actors are a pedestrian (moving to the right at speed one), a car (moving upwards at speed two), and the ego vehicle (moving upwards after waiting a bit).

Besides functions to initialize the scenario, module \lstinline+modele_scenario+ contains also a process with a parallel composition with an instance of process \lstinline+OBSTACLE+ for each moving obstacle and a process instantiating the ego vehicle as a parallel composition of the motion control process \lstinline+MOVE_CAR+ and the perception process \lstinline+LIDAR_MANAGER+.

Last, but not least, module \lstinline+modele_scenario+ contains also the processes \lstinline+SCHEDULER+ and \lstinline+RESTRAND+, which depend on the initial position of the car and the number of moving obstacles in the scenario.

\lstinputlisting{modele_scenario.lnt}

Different scenarios are obtained by simply modifying the contents of module \lstinline+modele_scenario+.

\subsection{Definitions for the environment of the autonomous vehicle}

Module \lstinline+modele_map_manager+ contains the process managing the ground truth map.
Observing the moves of the actors (car and obstacles), process \lstinline+MAP_MANAGER+ updates the map accordingly.
It also detects and signals a collision of the car with an obstacle.

Module \lstinline+modele_map_manager+ also contains the process \lstinline+ENVIRONMENT+, which models the environment of the autonomous vehicle, i.e., the parallel composition of the moving obstacles and the map manager.

\lstinputlisting{modele_map_manager.lnt}

\subsection{Definition of the principal process}

The main module contains the principal process \lstinline+MAIN+, mandatory to compile and generate the model. This process is the parallel composition of the two major processes \lstinline+ENVIRONMENT+ and \lstinline+CAR+, together with the processes \lstinline+SCHEDULER+ and \lstinline+RESTRAND+ adding the global constraints.

\begin{lstlisting}
module main_modele (modele_map_manager) is

------------------------------------------------------------------------
-- Main process
------------------------------------------------------------------------

process MAIN [GRID_UPDATE: G, OBSTACLE_POSITION: O, CAR_POSITION: C,
              GRID_CAR: CG, COLLISION, ARRIVAL: none, LIDAR_MAP: L,   
              END_OBSTACLE, TICK: none] is
  par CAR_POSITION, OBSTACLE_POSITION in
    par CAR_POSITION, GRID_CAR in
      CAR [LIDAR_MAP, CAR_POSITION, GRID_CAR, ARRIVAL]
    ||
      ENVIRONMENT [GRID_UPDATE, OBSTACLE_POSITION, CAR_POSITION, 
                   GRID_CAR, COLLISION, END_OBSTACLE]
    end par
  ||
    SCHEDULER [OBSTACLE_POSITION, CAR_POSITION, TICK]
  ||
    RESTRAND [OBSTACLE_POSITION, CAR_POSITION]
  end par
end process

------------------------------------------------------------------------

end module
\end{lstlisting}

\end{document}